\documentclass[graybox]{svmult}

\usepackage{type1cm}        
\usepackage{makeidx}         
\usepackage{graphicx}        
\usepackage{multicol}        
\usepackage[bottom]{footmisc}
\usepackage{makecell}
\usepackage{framed,multirow}
\usepackage{threeparttable}
\usepackage{pifont}
\usepackage{amssymb}
\usepackage{booktabs}
\usepackage{marvosym}
\usepackage{xcolor}
\usepackage{colortbl}
\usepackage[colorlinks,
            anchorcolor=red,
            citecolor=citecolor, 
            linkcolor=linkcolor,
            ]{hyperref}
\definecolor{citecolor}{HTML}{0071BC}
\definecolor{linkcolor}{HTML}{ED1C24}
\newcommand{\etal}{\mbox{et al.}}
\newcolumntype{P}[1]{>{\centering\arraybackslash}p{#1}}

\definecolor{igray}{rgb}{0.00, 0.00, 0.00}

\usepackage{newtxtext}       %
\usepackage[varvw]{newtxmath}       

\makeindex             

\begin{document}

\title*{Analyzing Tumors by Synthesis}
\author{Qi Chen, Yuxiang Lai, Xiaoxi Chen, Qixin Hu, Alan Yuille and Zongwei Zhou}
\institute{Qi Chen \at University of Chinese Academy of Sciences \at \email{chenqi24@ucas.ac.cn}
\and Yuxiang Lai \at Emory University \at \email{ylai76@emory.edu}
\and Xiaoxi Chen \at University of Illinois Urbana-Champaign \at \email{xiaoxic3@illinois.edu}
\and Qixin Hu \at University of Southern California \at \email{qixinhu@usc.edu}
\and Alan Yuille \at Johns Hopkins University \at \email{ayuille1@jhu.edu}
\and Zongwei Zhou (\Letter) \at Johns Hopkins University \at \email{zzhou82@jh.edu}
}

\maketitle

\abstract{
Computer-aided tumor detection has shown great potential in enhancing the interpretation of over 80 million CT scans performed annually in the United States. However, challenges arise due to the rarity of CT scans with tumors, especially early-stage tumors. Developing AI with real tumor data faces issues of scarcity, annotation difficulty, and low prevalence. Tumor synthesis addresses these challenges by generating numerous tumor examples in medical images, aiding AI training for tumor detection and segmentation. Successful synthesis requires realistic and generalizable synthetic tumors across various organs. This chapter reviews AI development on real and synthetic data and summarizes two key trends in synthetic data for cancer imaging research: modeling-based and learning-based approaches. Modeling-based methods, like Pixel2Cancer, simulate tumor development over time using generic rules, while learning-based methods, like DiffTumor, learn from a few annotated examples in one organ to generate synthetic tumors in others. Reader studies with expert radiologists show that synthetic tumors can be convincingly realistic. We also present case studies in the liver, pancreas, and kidneys reveal that AI trained on synthetic tumors can achieve performance comparable to, or better than, AI only trained on real data. Tumor synthesis holds significant promise for expanding datasets, enhancing AI reliability, improving tumor detection performance, and preserving patient privacy.
}

\section{Introduction}

Medical image analysis aims to derive detailed information non-invasively about a patient's medical condition, including the disease's origin, precise location, and its relationship with adjacent tissues. Specifically, for symptoms that cannot be directly diagnosed, medical professionals employ various imaging devices to capture detailed images of target organs for disease screening, diagnosis, and treatment. Therefore, medical imaging systems generate vast amounts of medical image data daily. This data may encompass different organs of the body, as well as tissues and pathological regions associated with diseases.

\begin{figure*}[t]
    \centering
    \includegraphics[width=1.0\linewidth]{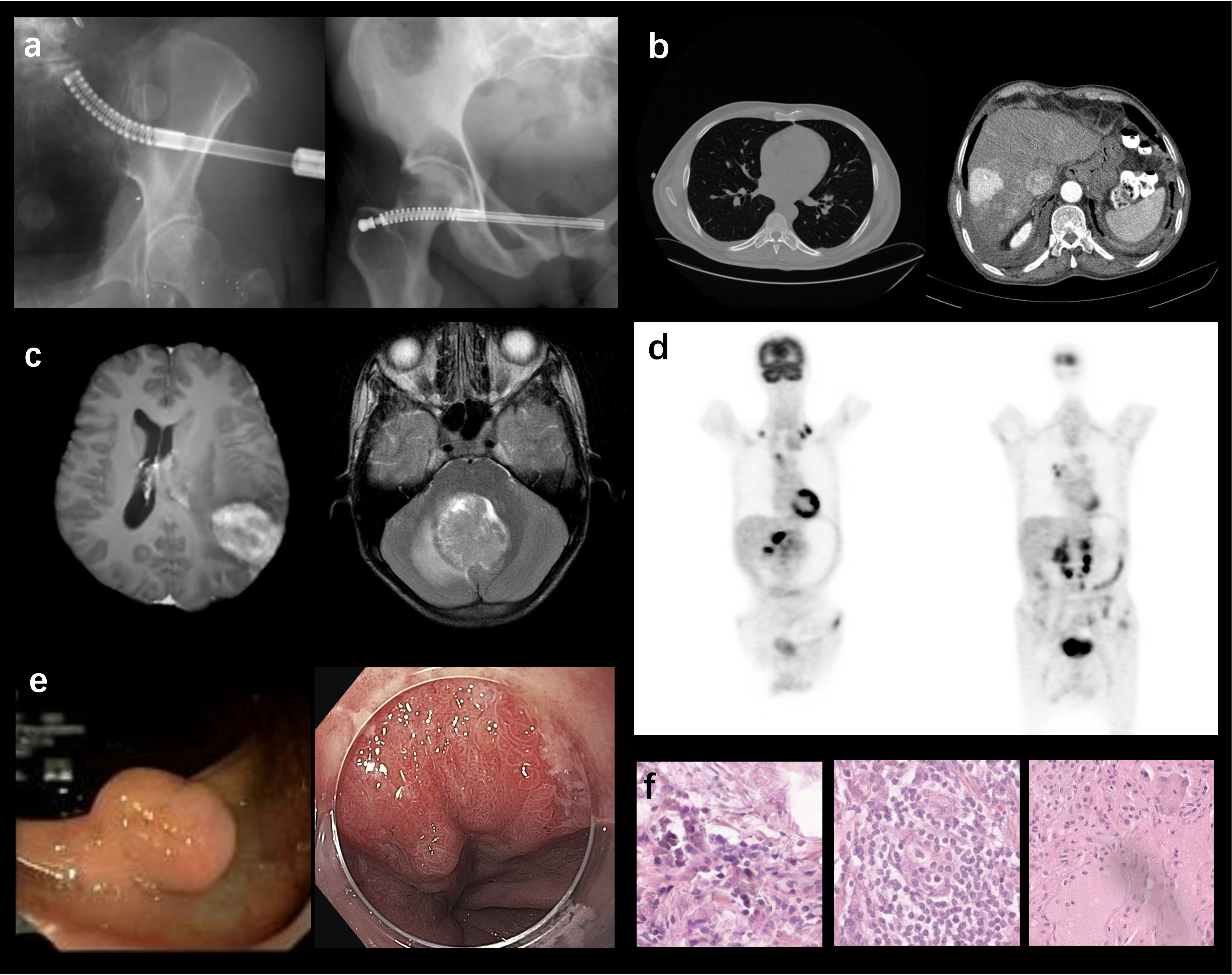}
    \caption{\textit{Can you distinguish synthetic data from real data in different modalities?} (a) X-ray image examples~\cite{gao2023synthetic}. (b) CT image examples~\cite{hamamci2023generatect}. (c) MR image examples~\cite{park2021generative}. (d) PET image example. (e) Endoscopy image example~\cite{li2024endora}. (f) Histopathology image example~\cite{aversa2024diffinfinite}.}
    \label{fig:real_synthetic_examples}
\end{figure*}

Medical images come from modalities like X-ray, CT, MRI, and PET. This data, referred to as \textbf{real data} in this chapter, can be analyzed using post-processing and artificial intelligence (AI) to reveal details not visible to the naked eye, aiding in disease detection such as detecting tumors at their early stage. However, managing real-world data for AI-driven diagnostics is challenging. \textbf{Synthetic data} offers a promising alternative, potentially allowing AI to generalize better to real-world scenarios and overcome the difficulties of using real data for training. Generally, synthetic data refer to artificially generated data that mimic the characteristics and structure of real data without being directly derived from actual observations (Fig.~\ref{fig:real_synthetic_examples}).

\subsection{Why Synthetic Data?}

Synthetic data are vital in AI research due to the challenges of acquiring real data, including time constraints, high costs, patient privacy concerns, and manual effort \cite{zhou2019models,feng2020parts2whole,haghighi2020learning,haghighi2021transferable}. They provide significant advantages by saving time and reducing the need for extensive manual annotation. The use of AI-generated content (AIGC) has proven effective across various domains, including medical imaging, where it serves both as a training resource for AI models and as a means of evaluating their performance with realistic yet hard-to-obtain data (see Table~\ref{tab:related_work_summary}). Synthetic data offer precise control over properties such as shape, texture, and location, which is particularly valuable in medical applications. This control enables the creation of diverse and representative datasets for model training and provides useful examples for medical education and patient communication. Additionally, controllable synthetic tumors facilitate AI debugging and model diagnostics, enhancing the interpretability of AI behavior. Increasing evidence supports that synthetic data can improve AI performance, making it a powerful tool for advancing research and improving outcomes in fields like oncology.

\begin{table*}[t]
    \centering
    \scriptsize
    \caption{
    A summary of existing synthesis methods for medical imaging encompasses various aspects: body part, disease type, imaging modality, dataset utilized, and generative model type. Additionally, we are maintaining a \href{https://github.com/MrGiovanni/SyntheticTumors/blob/main/AWESOME.md}{webpage} to track the latest publications and corresponding code repositories on data synthesis in medicine more comprehensively. 
    }\vspace{2px}
    \begin{tabular}{p{0.18\linewidth}p{0.11\linewidth}p{0.18\linewidth}P{0.1\linewidth}p{0.2\linewidth}p{0.18\linewidth}}
    \toprule
    reference & body part & disease & modality & dataset & generative model \\
    \midrule

    Teixeira~\etal~\cite{teixeira2018generating} & whole body & anomaly detection   & X-ray & private dataset   & GAN\\
    \rowcolor{igray!5}Wu~\etal~\cite{wu2018conditional} & chest & breast cancer   & X-ray & \makecell[tl]{DDSM~\cite{heath1998current}}   & GAN\\
    Gao~\etal~\cite{gao2023synthetic} & bone \& chest & COVID-19 lesion  & X-ray & \makecell[tl]{COVID-19 CXR~\cite{waheed2020covidgan}}   & GAN\\
    \rowcolor{igray!5}Jin~\etal~\cite{jin2018ct} & chest & lung nodule   & CT & \makecell[tl]{LIDC~\cite{armato2011lung}}   & GAN \\
    Yao~\etal~\cite{yao2021label} & chest & COVID-19   & CT & \makecell[tl]{LUNA16~\cite{setio2017validation}} &  hand-crafted designs  \\
    \rowcolor{igray!5}Jiang~\etal~\cite{jiang2020covid} & chest & Covid-19   & CT & \makecell[tl]{COVID~\cite{yang2020covid}} &  GAN  \\
    Jin~\etal~\cite{jin2021free} & abdomen & liver \& kidney tumors   & CT & \makecell[tl]{LiTS~\cite{bilic2023liver} \& KiTS~\cite{heller2021state}}   & GAN\\
    \rowcolor{igray!5}Wei~\etal~\cite{wei2022pancreatic} & abdomen & pancreatic tumors   & CT & private dataset   & GAN\\
    Lyu~\etal~\cite{lyu2022learning} & abdomen & liver tumors   & CT & \makecell[tl]{LiTS~\cite{bilic2023liver}}   & GAN\\
    \rowcolor{igray!5}Hu~\etal~\cite{hu2023label} & abdomen & liver tumors   & CT & \makecell[tl]{LiTS~\cite{bilic2023liver}} & hand-crafted designs\\
    Li~\etal~\cite{li2023early} & abdomen & pancreatic tumors   & CT & \makecell[tl]{MSD~\cite{antonelli2022medical}}  & hand-crafted designs \\
    \rowcolor{igray!5}Chen~\etal~\cite{chen2024towards} & abdomen & tumors in liver, pancreas and kidney  & CT & \makecell[tl]{MSD~\cite{antonelli2022medical} \& KiTS~\cite{heller2021state}} & Diffusion Model\\ 
    Lai~\etal~\cite{lai2024pixel} & abdomen & tumors in liver, pancreas and kidney  & CT & \makecell[tl]{MSD~\cite{antonelli2022medical} \& KiTS~\cite{heller2021state}} & hand-crafted designs\\ 
    \rowcolor{igray!5}Yu~\etal~\cite{yu20183d} & brain & brain tumors   & MRI & \makecell[tl]{BraTS \cite{menze2014multimodal}}  & GAN  \\
    Han~\etal~\cite{han2019infinite} & brain & brain tumors   & MRI & \makecell[tl]{BraTS~\cite{menze2014multimodal}}   & GAN\\
    \rowcolor{igray!5}Zhao~\etal~\cite{zhao2020tripartite} & abdomen & liver tumors   & MRI & private dataset   & GAN\\
    Mukherkjee~\etal~\cite{mukherkjee2022brain} & brain & brain tumors   & MRI & \makecell[tl]{BraTS~\cite{menze2014multimodal}}  & GAN\\
    \rowcolor{igray!5}Basaran~\etal~\cite{basaran2023lesionmix} & brain & brain tumors  & MRI   & \makecell[tl]{WMH \cite{kuijf2019standardized}}  & hand-crafted designs \\
    Wang~\etal~\cite{wang2018locality} & brain & -   & PET & private dataser  & GAN \\
    \rowcolor{igray!5}Luo~\etal~\cite{luo2022adaptive} & brain & -   & PET & private dataset  & GAN \\
    Sharan~\etal~\cite{sharan2021mutually} & mitral valve & landmark detection   & Endo. & \makecell[tl]{surgical simulator~\cite{engelhardt2019flexible}}  & GAN \\
    \rowcolor{igray!5}Yoon~\etal~\cite{yoon2022colonoscopic} & colon & polyp detection   & Endo. & private dataset  & GAN \\
    Hou~\etal~\cite{hou2019robust} & tissue & cancer & Histo. & \makecell[tl]{Kumar~\cite{kumar2017dataset}}  & GAN \\
    \rowcolor{igray!5}Xue~\etal~\cite{xue2021selective} & tissue & cancer   & Histo. & \makecell[tl]{PCam~\cite{veeling2018rotation}}  & GAN \\
    Aversa~\etal~\cite{aversa2024diffinfinite} & tissue & cancer & Histo. & private dataset  & Diffusion Model \\
    \rowcolor{igray!5}Du~\etal~\cite{du2023boosting} & skin & dermatoscopic lesion   & Dermo. & \makecell[tl]{ISIC~\cite{codella2018skin}}  & Diffusion Model \\
    \bottomrule
    \end{tabular}
    \begin{tablenotes}
        \item GAN stands for Generative Adversarial Network
        \item Endo. refers to Endoscopy, Histo. refers to Histopathology, and Dermo. refers to dermoscopy.
    \end{tablenotes}
    \label{tab:related_work_summary}
\end{table*}

\subsection{Real vs. Synthetic Data}

Unlike real data, which are collected from real-world imaging devices and represent true observations, synthetic data are created using algorithms, simulations, or models designed to replicate the properties of real data. We summarize the differences between real data and synthetic data as follows:
\textit{First}, real data are collected from actual imaging devices (e.g., X-ray, CT, MRI, PET, ultrasound, histopathology). Synthetic data are generated artificially using computational methods and simulations~\cite{hu2023label,lai2024pixel}.
\textit{Second}, real data are often limited by practical constraints such as patient privacy, the cost of imaging, and the time required for data collection and annotation \cite{chou2024acquiring}. Synthetic data can be produced in large quantities without the ethical and logistical issues associated with real data collection~\cite{chiruvella2021ethical}.
\textit{Third}, real data require manual annotation by experts, which is time-consuming \cite{park2020annotated} and prone to human error \cite{li2024abdomenatlas}. Annotations of synthetic data can be automatically generated as part of the data creation process, ensuring consistency and accuracy~\cite{qu2024abdomenatlas}.

\section{Detecting Real Tumors in CT Scans}

\begin{figure*}[t]
    \centering
    \includegraphics[width=1.0\linewidth]{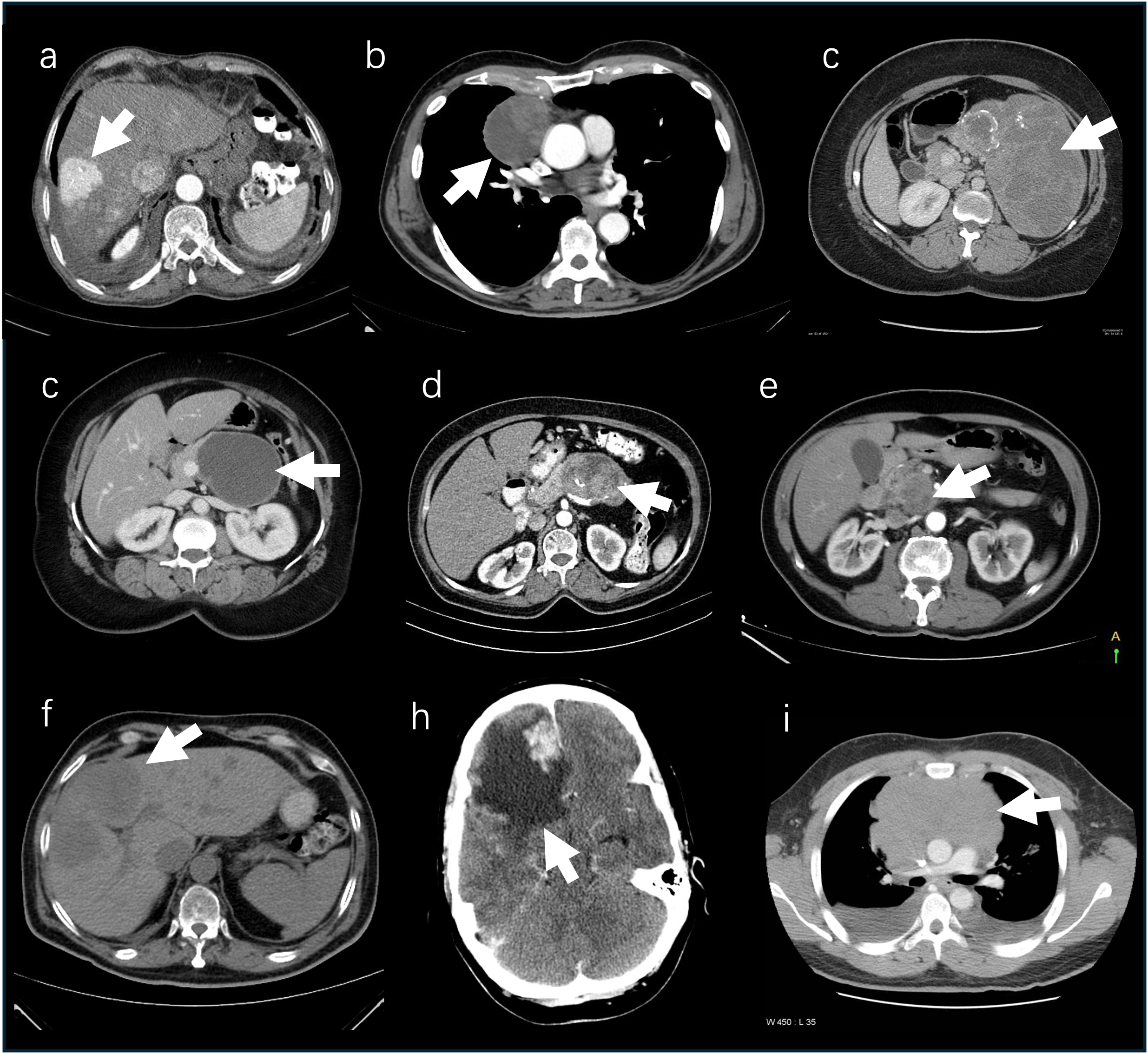}
    \caption{Tumors in solid organs. (a) Hepatocellular carcinoma. (b) Thymoma. (c) Solid pseudopapillary tumor of the pancreas. (d) Pancreatic mucinous cystadenoma. (e) Pancreatic mucinous cystadenocarcinoma. (f) Pancreatic adenocarcinoma. (g) Neuroendocrine tumor in liver. (h) Meningioma. (i) Mediastinal lymphoma.
}
    \label{fig:solid_organ_tumor}
\end{figure*}

\subsection{Tumors in Solid Organs}

Solid tumors in organs like the liver, kidneys, and brain---such as hepatocellular carcinoma, renal cell carcinoma, and glioma---typically show well-defined margins and growth patterns~\cite{sahani2016abdominal}, illustrated in Fig.~\ref{fig:solid_organ_tumor}. In CT images, early-stage tumors appear as small nodules with slightly blurred edges and homogeneous texture. As tumors advance, they grow larger, become irregular in shape, and exhibit significant mass effect and infiltrative growth~\cite{dunnick2016renal}. Advanced tumors may also show hemorrhage, necrosis, and fibrosis, leading to a heterogeneous appearance~\cite{silverman2012oncologic}.

\begin{itemize}
    
    \item \textbf{Liver tumors:} Hepatocellular carcinoma (HCC) is the most common malignant liver tumor. Early HCC presents as a small, well-differentiated nodule with a good prognosis and low metastatic potential~\cite{fowler2021pathologic}. CT imaging may show mass effect extending beyond the liver, displacement of blood vessels, intrahepatic venous thrombosis, and bile duct obstruction~\cite{reynolds2015infiltrative}. HCCs often exhibit intense arterial-phase enhancement due to neoangiogenesis and reduced portal triads, and typically appear hypoattenuating on venous phase scans~\cite{ayuso2018diagnosis}.

    \item \textbf{Pancreatic tumors:} Pancreatic ductal adenocarcinoma (PDAC) constitutes the majority of malignant pancreatic tumors and is associated with a very poor prognosis and high morbidity. In the early stages, PDACs typically appear as homogeneous small nodules with blurred edges. Secondary findings associated with advanced PDACs include contour abnormalities, abrupt termination of the biliary or pancreatic duct, pancreatic atrophy upstream from the mass, vascular encasement, etc.~\cite{laeseke2015combining}. PDACs typically exhibit poor enhancement, appearing hypoattenuating relative to the surrounding pancreatic parenchyma. This hypoenhancement is attributed to the development of a dense fibroblastic stromal component in PDACs~\cite{elbanna2020imaging}.

    \item \textbf{Kidney tumors:} Renal cell carcinoma (RCC) is the most common adult renal epithelial cancer, accounting for more than 90\% of all renal malignancies~\cite{leveridge2010imaging}. The most prevalent subtype, clear cell RCC, presents as a homogeneously enhancing lesion during the corticomedullary phase and as a hypoattenuating renal lesion surrounded by homogenously enhancing renal parenchyma in the nephrographic phase. Advanced clear cell RCC often appears heterogeneous in imaging due to the presence of hemorrhage, necrosis, and cysts, along with invasion into the renal pelvis, perirenal fat, or renal vessels~\cite{pmid29668296}. 
\end{itemize}

\subsection{Tumors in Tubular Organs}
\begin{figure*}[t]
    \centering
    \includegraphics[width=1.0\linewidth]{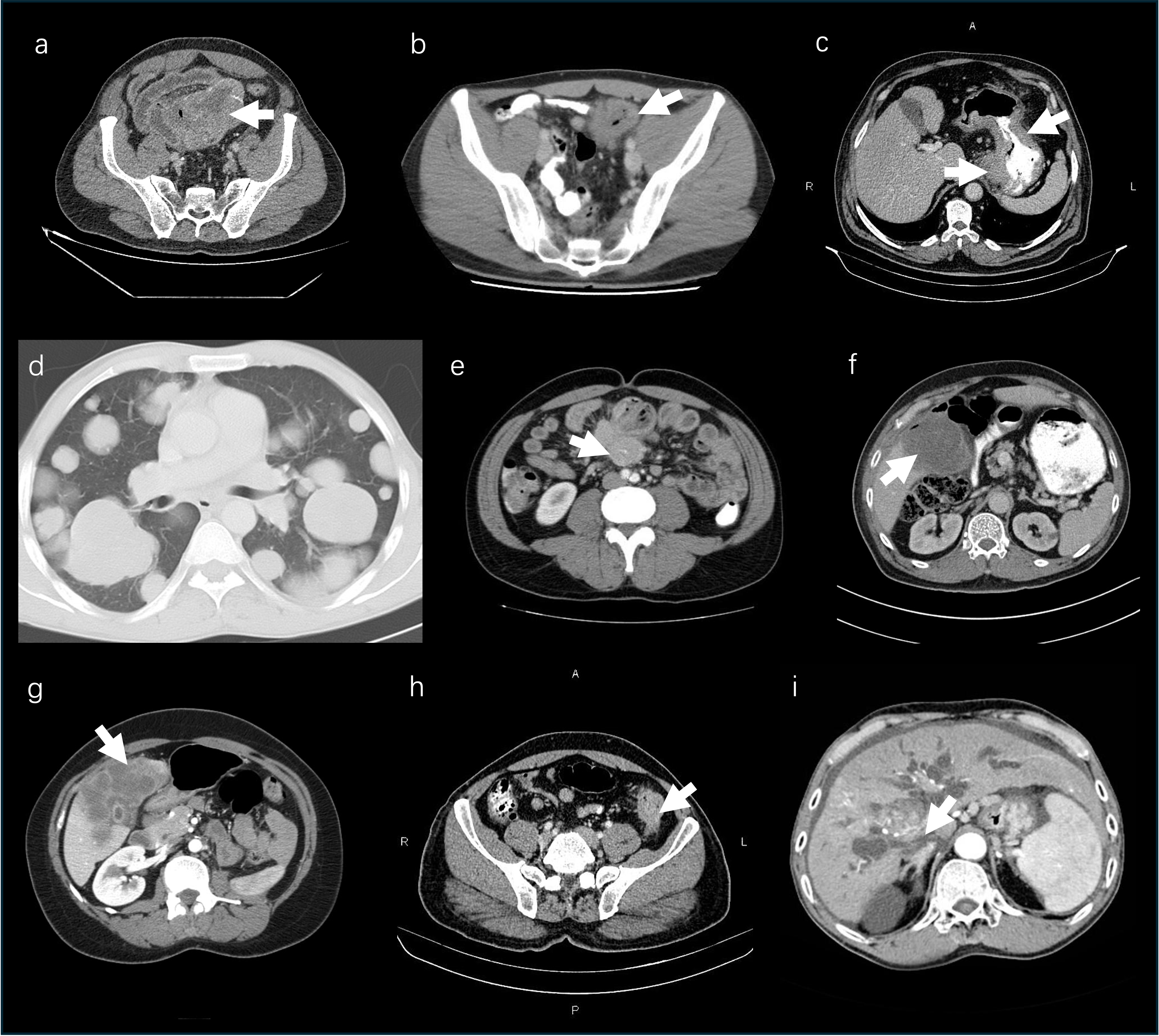}
    \caption{Tumors in tubular organs. (a) Gastrointestinal stromal tumor. (b) Sigmoid colon cancer. (c) Gastric cancer. (d) Lung metastases. (e) Intestinal carcinoid tumor. (f) Gallbladder carcinoma. (g) Gallbladder adenocarcinoma. (h) Colon cancer. (i) Cholangiocarcinoma.}
    \label{fig:tubular_organ_tumor}
\end{figure*}

Tumors in tubular organs, illustrated in Fig.~\ref{fig:tubular_organ_tumor}, have distinct growth patterns~\cite{nerad2016diagnostic}: exophytic, where the tumor expands into the lumen, and invasive, where it penetrates the organ wall into adjacent structures. For instance, colon cancer progresses from stage 0, confined to the lining, to stage I, invading the submucosa, and stage II, extending through the wall without nearby invasion~\cite{edge2010american}. Stage III involves spread to lymph nodes, and stage IV features metastasis to distant organs such as the liver or lungs. We also highlight unique characteristics of representative tubular tumors to illustrate their specific behaviors and progression.

\begin{itemize}

    \item \textbf{Esophageal tumors}, though rare, have a poor prognosis if malignant unless detected early and surgically removed~\cite{iyer2004imaging}. Imaging studies include X-ray esophagography, CT, endoscopic ultrasound, and PET. Malignant strictures often show asymmetric narrowing with abrupt margins and irregular, nodular, or ulcerated surfaces. X-ray esophagography helps evaluate invasion of the muscularis mucosae for early-stage cancers. Key CT features are eccentric or circumferential wall thickening over 5 mm and periesophageal soft tissue and fat stranding~\cite{lewin1996tumors,stout1957tumors}.

    \item \textbf{Stomach tumors}, primarily adenocarcinoma, are common and often asymptomatic when superficial. Up to 50\% of patients may have nonspecific gastrointestinal symptoms like dyspepsia. Endoscopy is the most sensitive method for diagnosis, allowing direct visualization and biopsy. Initial detection is often through radiological methods, with CT imaging using negative contrast to reveal common features such as polypoid masses, wall thickening, or ulceration~\cite{stout1953tumors,davis2000tumors,lewin1996tumors}.

    \item \textbf{Colorectal tumors} are a leading gastrointestinal malignancy. Contrast-enhanced CT of the chest, abdomen, and pelvis is used for staging, detecting metastases, evaluating surgical options, and assessing treatment response. Colorectal cancers typically appear as soft tissue masses that narrow the bowel lumen. Larger tumors often show ulceration, mucinous tumors may appear as low-density masses with low-density lymph nodes, and psammomatous calcifications can be seen in mucinous adenocarcinoma.~\cite{soga2005early,griswold1975colon}.
    
\end{itemize}

\begin{figure*}[t]
    \centering
\includegraphics[width=1.0\linewidth]{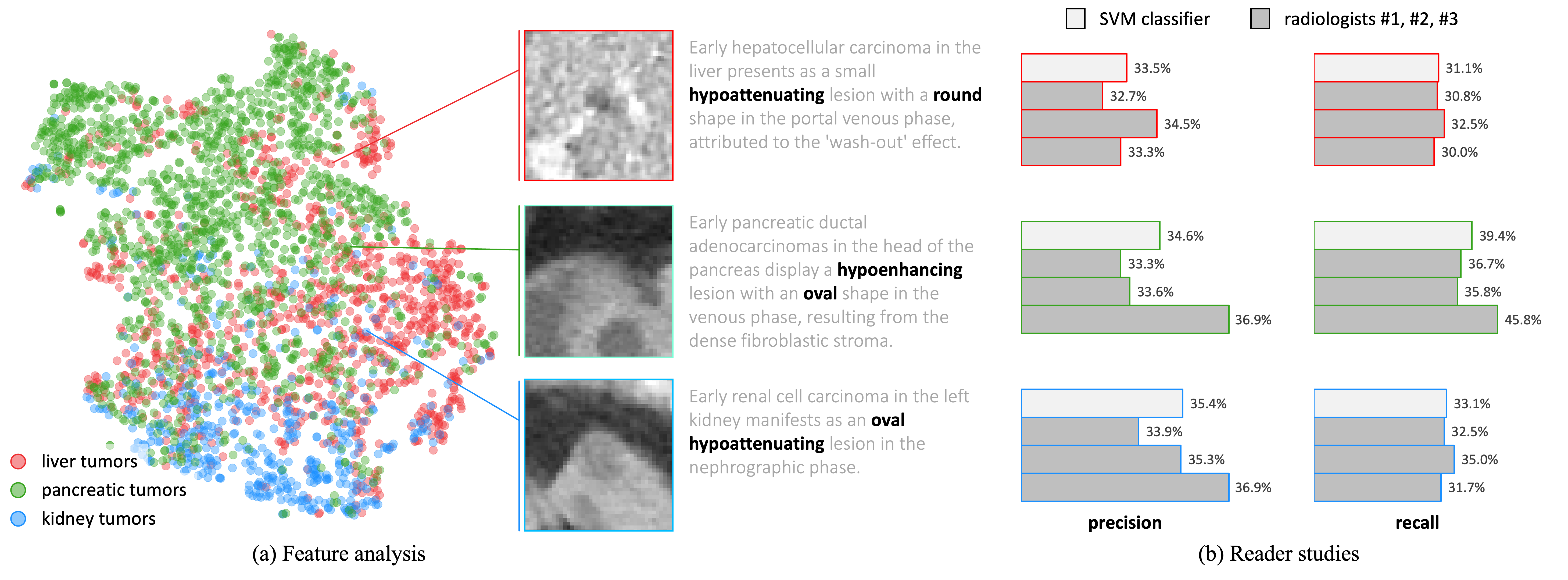}
    \caption{
    \textit{Feature analysis and reader study.} The left panel features a t-SNE (t-distributed stochastic neighbor embedding) visualization that maps the multidimensional Radiomics features of tumors from the liver, pancreas, and kidneys onto a two-dimensional space. This visualization underscores the substantial overlap in features among early-stage tumors from different organs, which may contribute to the challenges in correctly identifying their organ types. Complementing these findings, this study evaluates the efficacy of a support vector machine (SVM) classifier, which utilizes Radiomics Features~\cite{chu2019utility,wang2017comparison}, in differentiating the organ types for the cropped tumors. The SVM classifier is trained to classify each tumor as originating from either the liver, pancreas, or kidneys—a three-way classification challenge. Parallel to the assessment of the SVM classifier, three expert radiologists conducted a similar evaluation by reviewing the original CT scans containing these tumors. The results displayed on the right panel reveal significant difficulties faced by both the SVM classifier and the radiologists when it comes to accurately pinpointing the origin of early-stage tumors.  The precision and recall metrics for both the machine and human methods approximate the performance expected from random selection. 
    }
    \label{fig:preliminary}
\end{figure*}

\subsection{High Similarity in Early-Stage Tumors}

Early-stage tumors ($<$ 2cm) frequently exhibit similar imaging characteristics in CT scans, regardless of whether they originate in the liver, pancreas, or kidneys~\cite{choi2014ct}. Should this finding be validated, it could carry profound implications for the application of generative AI in medical imaging.  This implies that both modeling-based and learning-based approaches could be developed on a single tumor type with readily available annotated data and subsequently applied to synthesize various tumor types in other organs, for which data/annotation acquisition is more arduous. 

A study involving three expert radiologists was conducted to assess their ability to recognize the organ class of early-stage cancers~\cite{chen2024towards}. Three expert radiologists, certified in accordance with the Quality Standards Act, participated in the reader study. Recognition results are presented in Fig.~\ref{fig:preliminary}(b). The results were sufficiently compelling that medical professionals, boasting over five years of experience, could potentially mistake the synthetic tumors for genuine ones. The precision and recall scores, which approximate randomness, imply that the similarity in the appearance of early-stage tumors is such that even seasoned radiologists encounter difficulties when attempting to distinguish the organ types of these tumors.

The similarity of early-stage tumors can be evidenced by their Radiomics Feature\footnote{Utilization of the official radiomics feature repository~\cite{van2017computational} enables the extraction of appearance features, comprising 3D shape-based features, gray level co-occurrence matrix, gray level run length matrix, gray level size zone matrix, neighboring gray-tone difference matrix, and gray level dependence matrix.} profiles~\cite{chen2024towards}. From a \textit{qualitative} standpoint, Fig.~\ref{fig:preliminary}(a) depicts the feature mapping in a two-dimensional space via \textit{t}-SNE. The appearance features of early-stage tumors manifest within a joint feature space, with no discernible segregation among different organ types. From a \textit{quantitative} perspective, we trained a support vector machine (SVM) classifier to identify the organ types of early-stage tumors. To infer a general conclusion, we conducted ten repeated experiments and computed the precision and recall metrics of the SVM classifier for both the training and test sets. The final results indicate that both the precision and recall metrics for the training set are close to 1.0, demonstrating that the SVM is effectively trained and has established a robust decision boundary for the training set. However, the precision scores for the test set approximate random chance, as illustrated in Fig.~\ref{fig:preliminary}(b). This implies that even an effectively trained SVM classifier encounters difficulty in recognizing the organ types of unseen early-stage tumors. 

\section{Technical Barriers and Clinical Needs}

\subsection{Technical Barriers}

AI development for real tumors faces key technical barriers:
First, \textbf{data scarcity}: High-performance models need extensive annotated data, which is limited due to the time and expertise required for medical image and genomic annotation. Rare cancers further exacerbate this issue, leading to poor model performance on less common types \cite{zhou2021towards,zhou2022interpreting,tang2024efficient,chou2024embracing}.
Second, \textbf{generalization to different organs}: AI models struggle to generalize across organs due to distinct anatomical structures and imaging modalities. Models trained on one organ, like the lung, perform poorly on others, such as the liver, due to differing tissue compositions and imaging techniques \cite{zhang2023continual,liu2024universal}.
Third, \textbf{generalization to different demographics}: Privacy laws restrict access to diverse datasets, impacting model robustness. Variations in imaging protocols and genetic differences across populations can lead to biased models that perform poorly on underrepresented groups \cite{li2024well}.

\subsection{Clinical Needs}

Cancer research addresses critical clinical needs to improve patient outcomes and advance oncology. Key objectives include early cancer detection, developing effective treatments, and personalizing care strategies to enhance treatment success and system efficiency.

\smallskip\noindent\textbf{Early Detection and Diagnosis:}
Early detection and diagnosis of cancer are crucial for improving patient outcomes, as identifying cancer at an early stage often leads to more effective treatment and better survival rates. There is a critical need for screening methods with high sensitivity (ability to correctly identify those with cancer) and high specificity (ability to correctly identify those without cancer). Improved accuracy reduces false positives and false negatives, which are common issues in current screening practices. Besides, enhanced accuracy can help avoid overdiagnosis, where non-life-threatening cancers are treated unnecessarily, causing undue stress and potential harm to patients.

\smallskip\noindent\textbf{Health System Efficiency:} For effective training in tumor detection across multiple organs, AI models traditionally require numerous annotated real tumor examples from each organ~\cite{zhu2022assembling,kang2023label,qu2023annotating,liu2023clip}. However, these AI models often face difficulties in generalizing their ability to interpret images from different hospitals, a challenge compounded by varying imaging protocols, patient demographics, and scanner manufacturers~\cite{orbes2019multi,yan2020mri}. While the challenge of domain generalization could be alleviated if the AI is trained on a considerable number of annotated data from various domains \cite{xiao2022catenorm,yao2022unsupervised}, it could take up to 25 human years for just annotating tumors in a specific organ~\cite{xia2022felix,abi2023automatic}. Collecting and annotating a comprehensive dataset that includes tumor examples from several organs ($N$) and numerous hospitals ($M$) is a formidable task, denoted by the complexity ($N\times M$). We hypothesize that \textit{tumor synthesis could address this challenge by creating various tumor types across non-tumor images from multiple hospitals, even if only one type of tumor is available and annotated.} This approach can simplify the complexity from $N$\,$\times$\,$M$ to $1$\,$\times$\,$M$.

\smallskip\noindent\textbf{Personalized Treatment Planning} Tumors within the same type of cancer can vary significantly at the genetic and molecular levels. Personalized treatment planning requires comprehensive genomic profiling to identify specific mutations, gene expression patterns, and other molecular characteristics that drive an individual’s cancer. This information helps in selecting targeted therapies that are more likely to be effective for that particular tumor profile. Understanding the diversity of cancer cells within a tumor can inform treatment strategies that target multiple pathways and cell populations simultaneously.

\section{Technology Trend I: Modeling-based Approaches}

\begin{figure*}[t]
    \centering
    \includegraphics[width=1.0\linewidth]{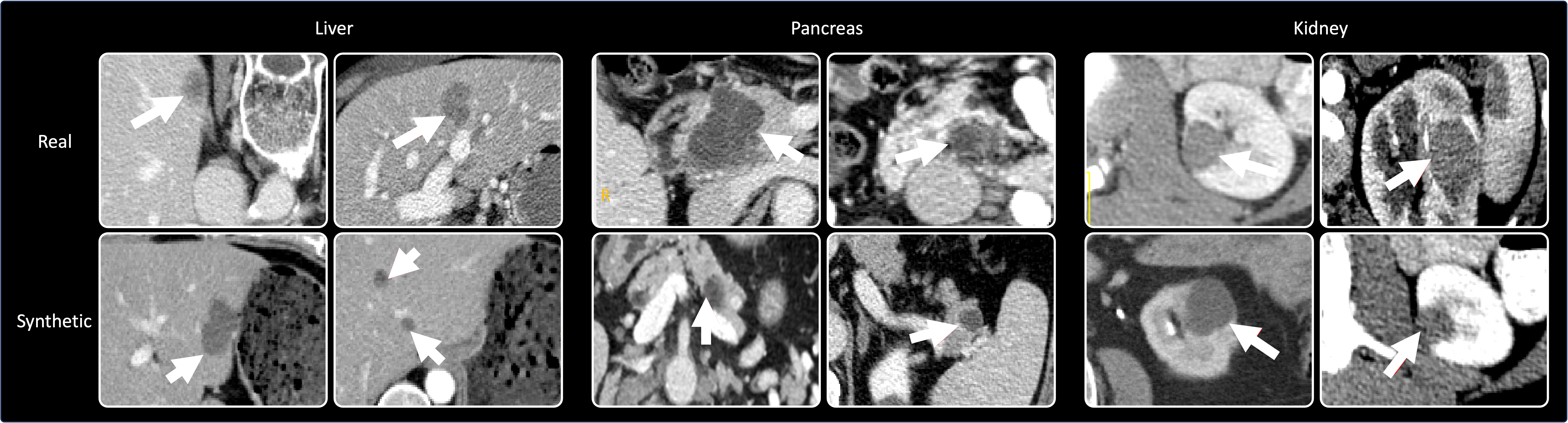}
    \caption{Synthetic liver, pancreatic, and kidney tumors generated by Cellular Automata \cite{lai2024pixel}.}
    \label{fig:CA_tumor_examples}
\end{figure*}

\textbf{Hand-crafted tumor synthesis} has been conducted in ~\cite{hu2022synthetic,hu2023label,hu2023synthetic,li2023early}. This approach applies a sequence of hand-crafted morphological image-processing operations, including local selection, texture generation, shape generation, and post-processing, to generate realistic tumors for training AI models. The intrinsic observation about these operations is clinical knowledge. Taking liver tumors as an example, the mean HU intensity of hepatocellular carcinomas (tumors grown from liver cells) was 106 HU (with a range of 36-162 HU)~\cite{lee2004triple}. Milder carcinomas usually lead to smaller, fewer spherical lesions, while multi-focal lesions (which means scattered small tumors) only appear in rare cases. Additionally, larger tumors usually display evident mass effects and are accompanied by capsule appearances that separate the tumor from the liver parenchyma~\cite{m2021use}. This medical guidance, together with visual clues, determines the parameters and pipeline of this method.

\begin{figure*}[t]
    \centering
\includegraphics[width=\linewidth]{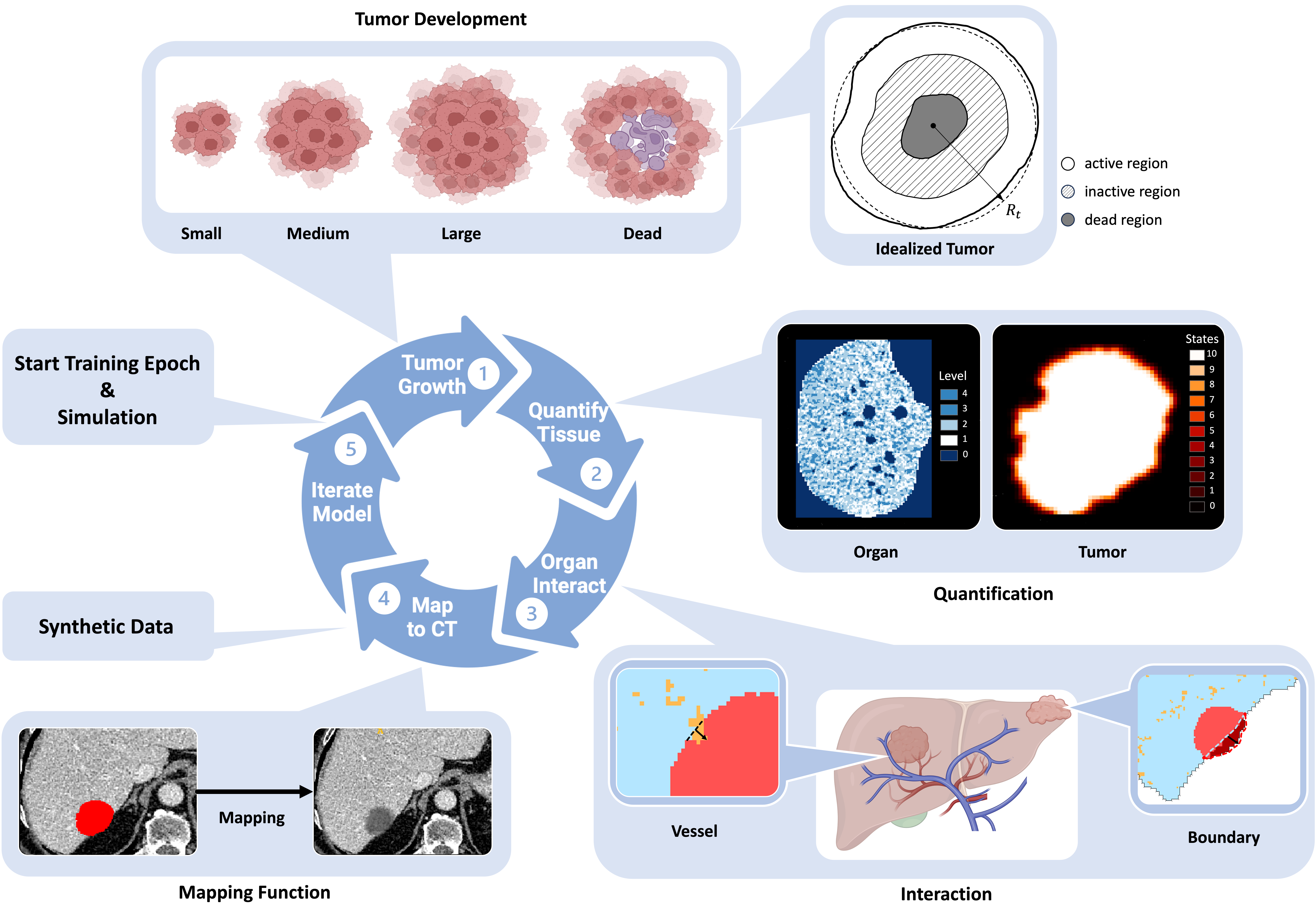} 
    \caption{
    \ding{172} Tumor development: The cellular automata simulate tumor growth from a single pixel to various sizes and even tumor death, producing synthetic tumors with diverse sizes, shapes, and textures. An idealized tumor is created to quantify development, with dead cells in the gray region, living quiescent cells in the inactive region, and proliferative cells in the active outer shell. \ding{173} Tissue quantification: The organ map in blue transforms CT images into distinct intensity levels affecting tumor development rate, while the red tumor map assigns values representing the tumor cell population. \ding{174} Tumor interaction with boundaries and vessels: The tumor grows and exerts pressure against organ boundaries and deforms as it interacts with vessels. \ding{175} Mapping synthetic tumors to CT images: A mapping function correlates the synthetic tumor with CT values, integrating the tumor's state with the original CT intensity. \ding{176} Training segmentation models with synthetic data: Pixel2Cancer generates new synthetic data for each epoch to train the segmentation model.
    }
    \label{fig:cellular_automata}
\end{figure*}

\textbf{Cellular Automata} are computational models used to simulate complex systems through simple rules and interactions. They employ a grid of cells (pixels), where each cell is initially assigned a state between zero and ten to represent the tumor population. The basic element is the \textit{cell}, which refers to a single \textit{pixel} in the computed tomography (CT) image. Tumor growth and behavior are modeled based on specific rules that simulate processes such as proliferation, invasion, and death. These rules are derived from medical knowledge and are guided by an idealized tumor model that reflects real-world characteristics. The tumor state can then be integrated into the original CT images to generate synthetic tumors in different organs. This tumor synthesis approach allows for sampling tumors at various stages and analyzing tumor-organ interactions. Motivated by this, Lai~\etal~\cite{lai2024pixel} proposed Pixel2Cancer to simulate tumor growth (Fig.~\ref{fig:cellular_automata}). Tumors generated by Pixel2Cancer are illustrated in Fig.~\ref{fig:CA_tumor_examples}.

\subsection{Property of Pixel2Cancer}
\smallskip\noindent\textbf{(\romannumeral1) Label-free.} Pixel2Cancer can be applied as a label-free data synthesis approach, eliminating the need for manual per-voxel annotation. Previous learning-based approaches, such as GANs~\cite{goodfellow2014generative} and Diffusion models~\cite{ho2020denoising}, are designed to learn the representation and distribution of tumors. While these approaches excel in generating natural images, synthesizing tumors in CT scans still requires significant amounts of paired tumor data. Moreover, when generating synthetic tumors, generative models also need masks to indicate the tumor locations and shapes~\cite{jin2021free}, necessitating extensive manual efforts for training and synthesis.

\smallskip\noindent\textbf{(\romannumeral2) Tumor development.} Pixel2Cancer incorporates specific medical knowledge regarding tumor growth and appearance, enabling the simulation of realistic tumors. None of the existing synthetic approaches can adequately simulate tumor development in abdominal CT, and the primary challenges in current synthetic methods are the proliferation and invasion of tumors~\cite{harpold2007evolution}. These processes in tumor growth are complex and interconnected, highly influenced by the surrounding environment~\cite{tanase2015complexity, wong2015pancreatic}. Consequently, synthetic tumors generated using current methods may conflict with normal organ structures and pose challenges when adapting them to different organs.

\smallskip\noindent\textbf{(\romannumeral3) Early tumor detection \& boundary segmentation.}
Early detection of small tumors is critical for timely cancer diagnosis. However, real datasets often lack sufficient instances due to the asymptomatic nature of early-stage patients. Pixel2Cancer can generate more small tumors to improve the sensitivity of segmentation models for small tumor detection. Additionally, Pixel2Cancer generates synthetic tumors with precise tumor masks, whereas real data annotations are often inaccurate at the boundaries, leading to \textit{label noise} in boundary segmentation accuracy.

\subsection{Clinical Perspectives}\label{sec:tumor-development
}

Tumors and genetic disorders from DNA mutations in single cells undergo complex growth processes~\cite{kumar2017robbins}. Mutations during cell division lead to uncontrolled proliferation, forming neoplastic lesions that can be benign or malignant~\cite{golias2004cell}. While both types follow similar growth principles, they differ in growth rate and invasiveness. Malignant tumors often grow rapidly, secreting growth factors or inducing surrounding stromal cells to do so, as seen in pancreatic IPMN lesions which grow larger and faster than benign ones~\cite{kang2011cyst}. Slow growth rates in renal tumors and hepatocellular carcinoma correlate with lower malignancy~\cite{smaldone2012small}. Malignant tumors are invasive, gradually penetrating and destroying surrounding tissues, whereas benign tumors remain confined to their original sites. Even slowly growing malignant tumors can infiltrate neighboring structures. Tumor necrosis, caused by rapid proliferation exceeding vascular supply~\cite{hiraoka2010tumour}, appears as irregular hypo-attenuating areas in CT images and serves as a poor prognostic indicator~\cite{fowler2021pathologic}. The \textbf{death} rule models this necrosis. A hybrid cellular automaton model is proposed to simulate tumor development from single cells to invasive tumors, capturing their continuous progression and interactions within the microenvironment.

\section{Technology Trend II: Learning-based Approaches}

\begin{figure*}[t]
    \centering
    \includegraphics[width=1.0\linewidth]{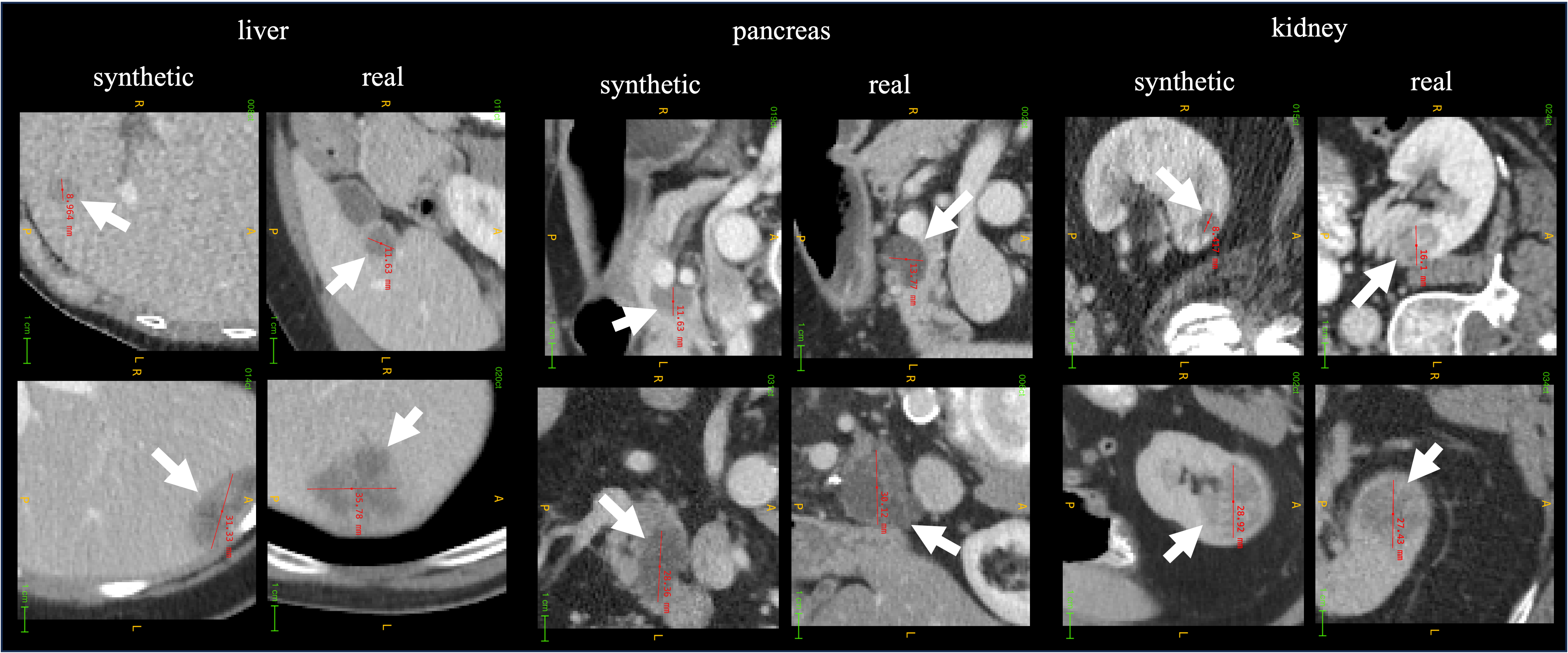}
    \caption{Examples of synthetic tumors generated by DiffTumor~\cite{chen2024towards} on the liver, pancreas, and kidney.}
    \label{fig:DiffTumor_tumor_examples}
\end{figure*}
 
\textbf{Generative Adversarial Networks} (GANs)~\cite{goodfellow2020generative} have been extensively explored for generating synthetic tumors. Zhao~\etal~\cite{zhao2020tripartite} introduced Tripartite-GAN, which simultaneously achieves contrast-enhanced magnetic resonance imaging synthesis and tumor detection. Mukherkjee~\etal~\cite{mukherkjee2022brain} proposed AGGrGAN to generate synthetic MRI scans of brain tumors. However, the training process of GANs is often unstable, making it challenging to achieve convergence. Additionally, GANs can suffer from issues such as mode collapse, where the model generates limited diversity in outputs.

\begin{figure*}[t]
    \centering
    \includegraphics[width=\linewidth]{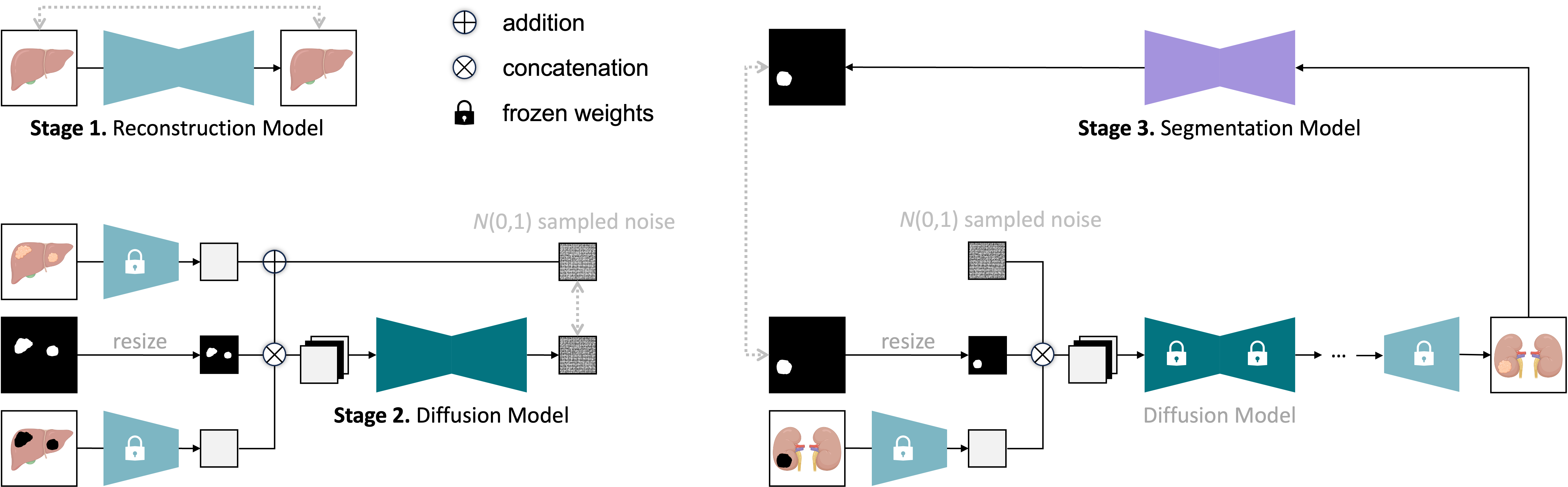}
    \caption{
    \textit{Overview of DiffTumor framework.} 
    In pursuit of achieving generalizable tumor synthesis, DiffTumor encompasses three stages. \ding{172} The first stage is the training of an Autoencoder Model—comprising an encoder and a decoder—to learn comprehensive latent features. The learning objective in this stage entails image reconstruction conducted on 9,262 unlabeled three-dimensional CT volumes. Both the trained encoder and decoder are integral to subsequent stages. \ding{173} The second stage involves training a Diffusion Model—a specialized generative model—by utilizing latent features and tumor masks as conditions. Once trained, the model is capable of generating the requisite latent features for the reconstruction of CT volumes with tumors, utilizing arbitrary masks. \ding{174} The third stage entails training a Segmentation Model with CT volumes of synthetic tumors, reconstructed by the decoder. Armed with a considerable repository of healthy CT volumes, DiffTumor has the capacity to generate an extensive collection of synthetic tumors, which vary in location, size, shape, texture, and intensity, thus contributing to the enhancement of AI models for tumor detection and segmentation.
    }
    \label{fig:DiffTumor}
\end{figure*}

\textbf{Diffusion Models}~\cite{ozbey2023unsupervised,khader2023denoising} provide more stable and reliable training compared to GANs, as they gradually denoise data, making the optimization process easier to control. Additionally, Diffusion Models are capable of generating high-quality, diverse samples with fewer issues related to mode collapse. DiffTumor~\cite{chen2024towards} is the first to explore tumor synthesis in abdominal organs using Diffusion Models and demonstrates an efficient method for achieving generalizable tumor synthesis (Fig.~\ref{fig:DiffTumor}). Tumors generated by DiffTumor are illustrated in Fig.~\ref{fig:DiffTumor_tumor_examples}.

\subsection{Property of DiffTumor}
\noindent\textbf{(\romannumeral1) Reduced annotations for Diffusion Model.} The quality of synthetic data produced by a generative model typically depends heavily on the quantity and diversity of the paired training data used during the training phase~\cite{chlap2021review,jaipuria2020deflating}. Nevertheless, the relationship between the number of annotated real tumors required for training the Diffusion Model and the performance of the Segmentation Model has not been extensively studied. DiffTumor found that the relationship between the amount of paired training data and the quality of synthetic data is not necessarily linear. Remarkably, it requires only one annotated tumor to train the Diffusion Model and to generate synthetic tumors for the subsequent training of the Segmentation Model. This finding is in contrast to the conventional wisdom in computer vision~\cite{ramesh2022hierarchical}, which often necessitates extensive datasets for training generative models. The results suggest that a smaller number of real tumors may suffice for training the Diffusion Model, particularly for early-stage tumors. Such a discovery could have profound implications for improving efficiency and reducing the costs associated with training generative models in the field of medical imaging.

\smallskip\noindent\textbf{(\romannumeral2) Accelerated tumor synthesis.}
The speed at which synthetic tumors are generated is critical for the practical use of synthetic data, particularly for accelerating the training of segmentation models. The synthesis speed of DiffTumor is significantly affected by the choice of timestep ($T$). An investigation into the influence of timestep on the segmentation performance of the Segmentation Model has been conducted. It is observed that using DDPM~\cite{ho2020denoising} with one-step sampling, DiffTumor cannot synthesize realistic textures for both organ and tumor, which negatively affects the training of the Segmentation Model. Conversely, by setting $T$ higher than 1, DiffTumor can produce more realistic textures, leading to an enhanced performance of the segmentation model. Taking into account the balance between performance and synthesis efficiency, DiffTumor selects a timestep of $T=4$ as the default setting for early tumor synthesis. This selection strikes a balance that allows for the generation of high-quality synthetic data while maintaining an acceptable level of efficiency.

\subsection{Clinical Perspectives}

This learning-based approach can be widely applied because of the similar growth dynamics observed in tumors across various types and locations. Tumorigenesis is a complex, multistage process involving cellular and histological transformations, from precancerous lesions to malignant tumors. This progression, driven by genetic mutations and functional changes known as the `hallmarks of cancer,' is consistent across tumor types~\cite{lahouel2020revisiting, choi2014ct, hanahan2000hallmarks}.
Early-stage tumors typically consist of well-to-moderately differentiated cells with mild atypia and invasiveness, showing rare hemorrhage and necrosis. They often appear as homogeneous nodules with slightly indistinct margins and small diameters in CT images~\cite{chu2017diagnosis, skarin2015atlas}. In contrast, advanced tumors exhibit significant infiltration and destruction of surrounding tissues, extending beyond the original site and potentially affecting adjacent structures~\cite{ayuso2018diagnosis}. As tumors become more malignant, rapid growth leads to ischemia and necrosis due to insufficient vascular supply, resulting in heterogeneous patterns in CT images with features like hemorrhage and fibrosis~\cite{yee2021tumor, saar2008radiological}. These characteristics are consistent across different populations, ages, and genders.

\section{Tumor Synthesis Benchmark}
We evaluate the effectiveness of Pixel2Cancer and DiffTumor by comparing them with supervised models trained on real data and several prominent unsupervised anomaly segmentation methods. Table~\ref{tab:synt_real_result} highlights that synthetic data have significantly outperformed all these baseline methods, achieving a DSC of 59.77\% and NSD of 61.29\%. These results highlight the potential of synthetic strategies to avoid per-pixel manual annotation for tumor segmentation.

\begin{table*}[t]
    \centering
    \scriptsize
    \caption{
        \textit{Comparison with state-of-the-art unsupervised methods.} We compare the initial label-free modeling-based methods with other unsupervised anomaly segmentation baselines, tumor synthesis strategies, and fully-supervised methods. Modeling-based methods significantly outperform all other state-of-the-art unsupervised baseline methods and even surpass the fully-supervised method with detailed \textit{pixel-wise annotation}.
    }
    \begin{tabular}{p{0.1\linewidth}p{0.2\linewidth}p{0.18\linewidth}P{0.22\linewidth}P{0.12\linewidth}P{0.12\linewidth}}
    \multicolumn{5}{l}{\textit{Liver tumor segmentation performance.}} \\
        \hline
        tumors & method & architecture & labeled / unlabeled CTs & DSC (\%) & NSD (\%) \\
        \hline
        none & PatchCore~\cite{roth2022towards} & Resnet50 & 0 / 116 & 15.9  & 16.4 
        \\
        none & f-AnoGAN~\cite{schlegl2019f}     & Customized~\cite{baur2021autoencoders} & 0 / 116 & 19.0  & 16.9  
        \\
        none & VAE~\cite{kingma2013auto}        & Customized~\cite{baur2021autoencoders} & 0 / 116 & 24.6  & 23.6  
        \\
        synt & Yao~\etal~\cite{yao2021label} & U-Net & 0 / 116 & 32.8 & 31.3 
        \\
        \hline
        real & fully-supervised & U-Net & 101 / 0  & 56.7  & 58.0  
        \\ 
        synt & hand-crafted~\cite{hu2023label} & U-Net & 0 / 116 & 59.8 &  61.3
        \\
        synt & Pixel2Cancer~\cite{lai2024pixel} & U-Net & 0 / 116 & 58.9 &  63.7
        \\
        synt & DiffTumor~\cite{chen2024towards} & U-Net & 101 / 116 & 70.9 &  71.2
        \\
        \hline
        \\
    \end{tabular}
    \begin{tabular}{p{0.1\linewidth}p{0.2\linewidth}p{0.18\linewidth}P{0.22\linewidth}P{0.12\linewidth}P{0.12\linewidth}}
    \multicolumn{5}{l}{\textit{Pancreas tumor segmentation performance.}} \\
        \hline
        tumors & method & architecture & labeled / unlabeled CTs & DSC (\%) & NSD (\%) \\
        \hline
        real & fully-supervised & U-Net & 96 / 0  & 57.5  & 56.5  
        \\ 
        synt & hand-crafted~\cite{hu2023label} & U-Net & 0 / 104 & 54.1 &  52.2
        \\
        synt & Pixel2Cancer~\cite{lai2024pixel} & U-Net & 0 / 104 & 60.9 &  57.1
        \\
        synt & DiffTumor~\cite{chen2024towards} & U-Net & 96 / 104 & 64.8 &  60.5
        \\
        \hline
        \\
    \end{tabular} 
    \begin{tabular}{p{0.1\linewidth}p{0.2\linewidth}p{0.18\linewidth}P{0.22\linewidth}P{0.12\linewidth}P{0.12\linewidth}}
    \multicolumn{5}{l}{\textit{Kidney tumor segmentation performance.}} \\
        \hline
        tumors & method & architecture & labeled / unlabeled CTs & DSC (\%) & NSD (\%) \\
        \hline
        real & fully-supervised & U-Net & 96 / 0  & 71.3  & 62.8  
        \\ 
        synt & hand-crafted~\cite{hu2023label} & U-Net & 0 / 120 & 63.2 &  55.4
        \\
        synt & Pixel2Cancer~\cite{lai2024pixel} & U-Net & 0 / 120 & 73.2 &  65.0
        \\
        synt & DiffTumor~\cite{chen2024towards} & U-Net & 96 / 120 & 84.2 &  76.6
        \\
        \hline
    \end{tabular}
    \label{tab:synt_real_result}
\end{table*}

\subsection{Case Study: Fake Tumors, Real Results}

\noindent\textbf{Synthetic Liver Tumors.} In synthetic liver tumors, synthetic data have demonstrated significant superiority over real data across all stages, from small to large tumors. The hand-crafted approach proposed by~\cite{hu2023label} outperforms real data, achieving a $2.3\%$ improvement in DSC and a $3.3\%$ improvement in NSD. Pixel2Cancer has shown superior performance in liver segmentation, with a DSC improvement of $2.2\%$ and an NSD improvement of $5.7\%$. Additionally, DiffTumor has surpassed the performance of real data by $4.0\%$ in DSC and $4.7\%$ in NSD.

\noindent\textbf{Synthetic Kidney Tumors.}
In kidney tumors, segmentation models have achieved superior performance when using synthetic data as data augmentation. Pixel2Cancer has shown superior performance in kidney segmentation, with a DSC improvement of $2.4\%$ and an NSD improvement of $3.2\%$. Additionally, DiffTumor has surpassed the performance of real data by $7.0\%$ in DSC and $6.7\%$ in NSD.

\noindent\textbf{Synthetic Pancreatic Tumors.}
Segmentation models also have achieved superior performance in pancreatic tumor segmentation when using synthetic data as data augmentation. Pixel2Cancer has shown superior performance in pancreas segmentation, with a DSC improvement of $3.9\%$ and an NSD improvement of $1.9\%$. DiffTumor has surpassed the performance of real data by $8.2\%$ in DSC and $9.4\%$ in NSD.

\begin{figure*}[t]
    \centering
    \includegraphics[width=1.0\linewidth]{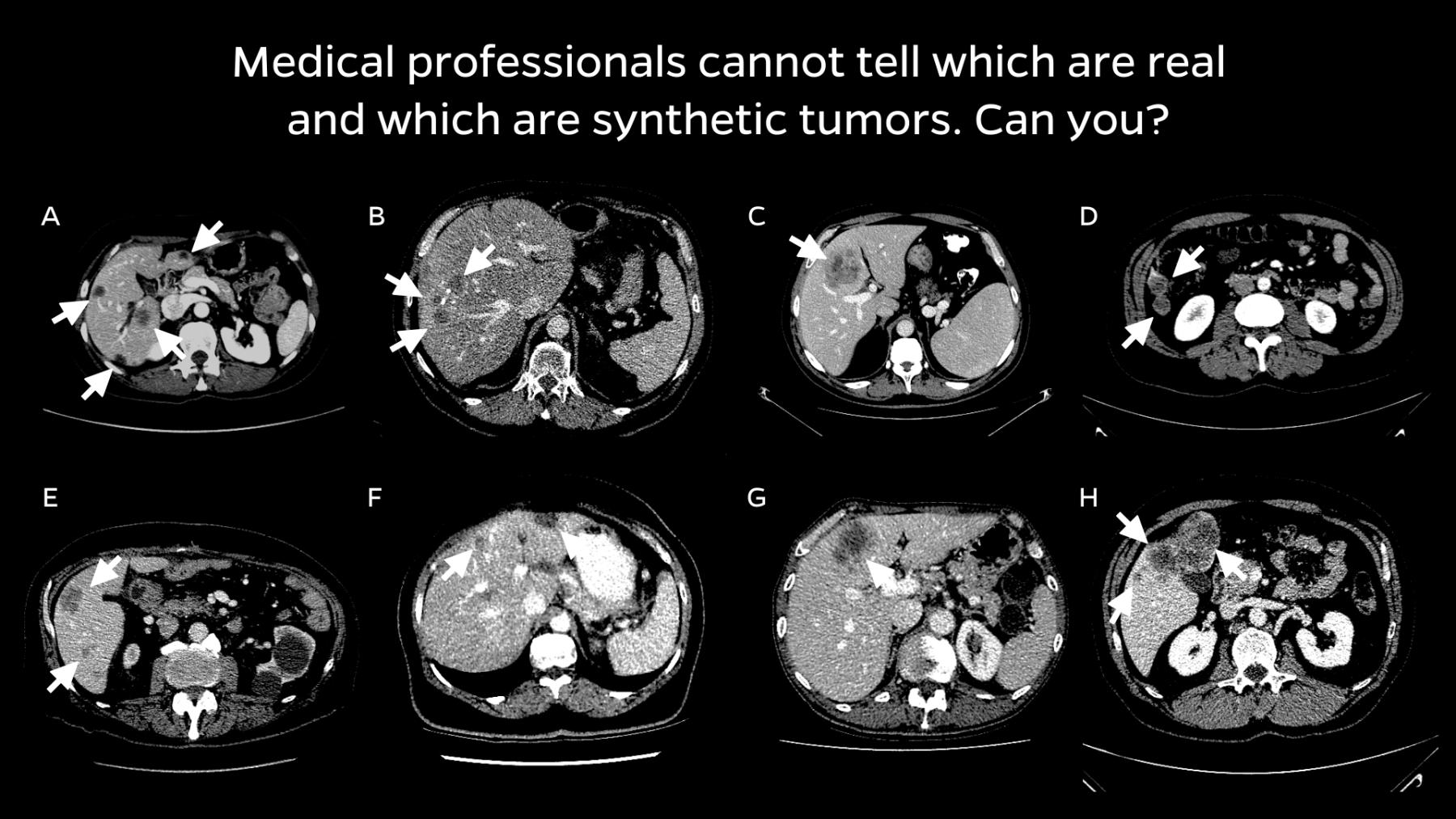}
    \caption{Visual Turing Test. Can you find some examples of synthetic data in the CT images? Answers are in \href{https://hub.jhu.edu/2024/05/30/researchers-create-artificial-tumors-for-cancer-research/}{\textit{Johns Hopkins Researchers Create Artificial Tumors to Help AI Detect Early-Stage Cancer}}.}
    \label{fig:visual_turing_test}
\end{figure*}

\begin{table}[h]
    \centering
    \scriptsize
    \caption{
    \textit{Results of reader study.} Pixel2Cancer\ (top table): The test was conducted with three medical professionals having 7, 9, and 14 years of experience, respectively. Each professional evaluated 50 CT images for each organ, consisting of both real and synthetic tumors. They were tasked with categorizing each CT image as either \textit{real}, \textit{synthetic}, or \textit{unsure}. DiffTumor (bottom table): Visual Turing test over three organs has been conducted with two radiologists (R1 and R2). Both radiologists are provided with 60 three-dimensional CT volumes of each organ, including 30 scans with real tumors and the remaining 30 with synthetic ones. Radiologists are tasked to label each CT volume as \textit{real} or \textit{synthetic}. A lower specificity score indicates a higher number of synthetic tumors being identified as real.
    }\vspace{2px}
    \begin{tabular}{p{0.1\linewidth}p{0.2\linewidth}P{0.22\linewidth}P{0.22\linewidth}P{0.22\linewidth}}
    \multicolumn{5}{l}{\textit{Modeling-based Approach: Pixel2Cancer}} \\
    \toprule
    reader & metric & liver & pancreas & kidneys \\
    \midrule
    \multirow{3}{*}{\makecell[l]{R1}} & sensitivity (\%) &100 &95.0 & 95.5\\
     & specificity (\%)&27.3 & 22.7 & 26.7\\
     & accuracy (\%) & 60.9 & 57.1& 67.6\\
    \midrule
    \multirow{3}{*}{\makecell[l]{R2}} & sensitivity (\%)&94.7 & 87.5 & 90.0\\
     & specificity (\%) &47.8 & 47.4 &56.3\\
     & accuracy (\%) &69.1 & 65.7 & 75.0\\
     \midrule
    \multirow{3}{*}{\makecell[l]{R3}} & sensitivity (\%) & 100 & 100 & 100\\
     & specificity (\%)& 45.4 & 55.6 & 57.9\\
     & accuracy (\%) & 68.4 & 72.4 & 75.8\\
    \bottomrule
    \end{tabular}
    \begin{tablenotes}
    \scriptsize
        \item positives: real tumors ($N$ = 25); negatives: synthetic tumors ($N$ = 25).
    \end{tablenotes}
    \begin{tabular}{p{0.1\linewidth}p{0.2\linewidth}P{0.22\linewidth}P{0.22\linewidth}P{0.22\linewidth}}
    \multicolumn{5}{l}{} \\
    \multicolumn{5}{l}{\textit{Learning-based Approach: DiffTumor}} \\
    \toprule
    reader & metric & liver & pancreas & kidneys \\
    \midrule
    \multirow{3}{*}{\makecell[l]{R1}} & sensitivity (\%) &100&97.1&92.9\\
     & specificity (\%) &2.9&0.0&3.1\\
     & accuracy (\%) &45.0 &56.7&45.0\\
    \midrule
    \multirow{3}{*}{\makecell[l]{R2}} & sensitivity (\%) &84.6&100&100\\
     & specificity (\%) &47.1&44.0&65.6\\
     & accuracy (\%) &63.3&76.7&81.7\\
    \bottomrule
    \end{tabular}
    \begin{tablenotes}
        \item positives: real tumors ($N$ = 30); negatives: synthetic tumors ($N$ = 30).
    \end{tablenotes}
    \label{tab:reader_studies}
\end{table}

\subsection{Visual Turing Test}\label{sec:visual_turing_test}

The Turing Test, introduced by Alan Turing in ``Computing Machinery and Intelligence'' \cite{turing2009computing}, assesses whether a machine can exhibit intelligent behavior indistinguishable from that of a human. We apply the Visual Turing Test to evaluate if synthetic tumors resemble real tumors. For this, we compared CT volumes containing real and synthetic tumors across various organs. Professionals, blinded to the origins of the samples, classified each volume as real or synthetic based on 3D views of continuous slice sequences (see Fig.~\ref{fig:visual_turing_test}). 

\noindent\textbf{Modeling-based approach (Pixel2Cancer).} The outcome metrics, as presented in Table~\ref{tab:reader_studies}, unveil the performance evaluations by different radiologists. For the junior radiologist R1 (7 years of experience), metrics such as accuracy, sensitivity, and specificity all register below 40\%. Notably, a specificity of 35.5\% indicates that 64.5\% of synthetic tumors are inaccurately identified as real. The intermediate radiologist R2 (9 years of experience) exhibits comparable metrics around 40\%, with 59.2\% of synthetic tumors causing confusion. Even the senior radiologist R3 (14 years of experience) misclassifies 44.1\% of synthetic tumors as real, underscoring the formidable challenge posed even to seasoned professionals. These results emphasize the efficacy of our modeling-based approach (Pixel2Cancer) in achieving a realistic simulation of tumor development.

\noindent\textbf{Learning-based approach (DiffTumor).} Two professionals were involved in this test, with 4 and 11 years of experience, respectively. 60 CT volumes were arranged in random order and scrutinized independently by two professionals. The test outcomes are detailed in Table~\ref{tab:reader_studies}.
Radiologists R1's near-zero specificity scores suggested that synthetic data closely resembled real tumors, resulting in the misclassification of most synthetic tumors as real. Consequently, R1's accuracy scores hovered around 50\%. Conversely, R2, with more experience, exhibited higher specificity scores compared to R1, approaching 50\%. This implies that nearly 50\% of synthetic samples were correctly identified as synthetic, indicating a better discernment between real and synthetic tumors by R2. These findings affirm the effectiveness of DiffTumor in generating visually realistic tumors.

\section{Conclusion}
In this chapter, we delved into the concept of synthetic data and its application in medical fields, with a particular focus on cancer research. We defined synthetic data and discussed its critical role in cancer research, such as improving data diversity, protecting patient privacy, and enabling robust research in tumor detection, diagnosis, and treatment. We explored the challenges and opportunities presented in cancer research. Despite these challenges, synthetic data offers significant opportunities, such as enhancing the training of machine learning models, supporting large-scale studies without privacy concerns, and fostering innovation in personalized medicine. Additionally, we highlighted promising approaches and future directions for the use of synthetic data in cancer research, including modeling-based methods and learning-based methods. These methods are opening new avenues for more accurate and comprehensive cancer research, enabling researchers to simulate various scenarios and treatments, and ultimately contributing to better patient outcomes.

\medskip\noindent\textbf{Acknowledgement.} This work was supported by the Lustgarten Foundation for Pancreatic Cancer Research and the Patrick J. McGovern Foundation Award.

\bibliographystyle{styles/spmpsci.bst}
\bibliography{refs,zzhou}

\begin{thebibliography}{100}
\providecommand{\url}[1]{{#1}}
\providecommand{\urlprefix}{URL }
\expandafter\ifx\csname urlstyle\endcsname\relax
  \providecommand{\doi}[1]{DOI~\discretionary{}{}{}#1}\else
  \providecommand{\doi}{DOI~\discretionary{}{}{}\begingroup \urlstyle{rm}\Url}\fi

\bibitem{abi2023automatic}
Abi~Nader, C., Vetil, R., Wood, L.K., Rohe, M.M., B{\^o}ne, A., Karteszi, H., Vullierme, M.P.: Automatic detection of pancreatic lesions and main pancreatic duct dilatation on portal venous ct scans using deep learning.
\newblock Investigative Radiology  (2023)

\bibitem{antonelli2022medical}
Antonelli, M., Reinke, A., Bakas, S., Farahani, K., Kopp-Schneider, A., Landman, B.A., Litjens, G., Menze, B., Ronneberger, O., Summers, R.M., et~al.: The medical segmentation decathlon.
\newblock Nature communications \textbf{13}(1), 1--13 (2022)

\bibitem{armato2011lung}
Armato~III, S.G., McLennan, G., Bidaut, L., McNitt-Gray, M.F., Meyer, C.R., Reeves, A.P., Zhao, B., Aberle, D.R., Henschke, C.I., Hoffman, E.A., et~al.: The lung image database consortium (lidc) and image database resource initiative (idri): a completed reference database of lung nodules on ct scans.
\newblock Medical physics \textbf{38}(2), 915--931 (2011)

\bibitem{aversa2024diffinfinite}
Aversa, M., Nobis, G., H{\"a}gele, M., Standvoss, K., Chirica, M., Murray-Smith, R., Alaa, A.M., Ruff, L., Ivanova, D., Samek, W., et~al.: Diffinfinite: Large mask-image synthesis via parallel random patch diffusion in histopathology.
\newblock Advances in Neural Information Processing Systems \textbf{36} (2024)

\bibitem{ayuso2018diagnosis}
Ayuso, C., Rimola, J., Vilana, R., Burrel, M., Darnell, A., Garc{\'\i}a-Criado, {\'A}., Bianchi, L., Belmonte, E., Caparroz, C., Barrufet, M., et~al.: Diagnosis and staging of hepatocellular carcinoma (hcc): current guidelines.
\newblock European journal of radiology \textbf{101}, 72--81 (2018)

\bibitem{basaran2023lesionmix}
Basaran, B.D., Zhang, W., Qiao, M., Kainz, B., Matthews, P.M., Bai, W.: Lesionmix: A lesion-level data augmentation method for medical image segmentation.
\newblock In: International Conference on Medical Image Computing and Computer-Assisted Intervention, pp. 73--83. Springer (2023)

\bibitem{baur2021autoencoders}
Baur, C., Denner, S., Wiestler, B., Navab, N., Albarqouni, S.: Autoencoders for unsupervised anomaly segmentation in brain mr images: a comparative study.
\newblock Medical Image Analysis \textbf{69}, 101952 (2021)

\bibitem{bilic2023liver}
Bilic, P., Christ, P., Li, H.B., Vorontsov, E., Ben-Cohen, A., Kaissis, G., Szeskin, A., Jacobs, C., Mamani, G.E.H., Chartrand, G., et~al.: The liver tumor segmentation benchmark (lits).
\newblock Medical Image Analysis \textbf{84}, 102680 (2023)

\bibitem{chen2024towards}
Chen, Q., Chen, X., Song, H., Xiong, Z., Yuille, A., Wei, C., Zhou, Z.: Towards generalizable tumor synthesis.
\newblock In: IEEE/CVF Conference on Computer Vision and Pattern Recognition (2024).
\newblock \urlprefix\url{https://github.com/MrGiovanni/DiffTumor}

\bibitem{chiruvella2021ethical}
Chiruvella, V., Guddati, A.K., et~al.: Ethical issues in patient data ownership.
\newblock Interactive journal of medical research \textbf{10}(2), e22269 (2021)

\bibitem{chlap2021review}
Chlap, P., Min, H., Vandenberg, N., Dowling, J., Holloway, L., Haworth, A.: A review of medical image data augmentation techniques for deep learning applications.
\newblock Journal of Medical Imaging and Radiation Oncology \textbf{65}(5), 545--563 (2021)

\bibitem{choi2014ct}
Choi, J.Y., Lee, J.M., Sirlin, C.B.: Ct and mr imaging diagnosis and staging of hepatocellular carcinoma: part i. development, growth, and spread: key pathologic and imaging aspects.
\newblock Radiology \textbf{272}(3), 635--654 (2014)

\bibitem{chou2024acquiring}
Chou, Y.C., Li, B., Fan, D.P., Yuille, A., Zhou, Z.: Acquiring weak annotations for tumor localization in temporal and volumetric data.
\newblock Machine Intelligence Research pp. 1--13 (2024).
\newblock \urlprefix\url{https://github.com/johnson111788/Drag-Drop}

\bibitem{chou2024embracing}
Chou, Y.C., Zhou, Z., Yuille, A.: Embracing massive medical data.
\newblock arXiv preprint arXiv:2407.04687  (2024).
\newblock \urlprefix\url{https://github.com/MrGiovanni/OnlineLearning}

\bibitem{chu2017diagnosis}
Chu, L.C., Goggins, M.G., Fishman, E.K.: Diagnosis and detection of pancreatic cancer.
\newblock The Cancer Journal \textbf{23}(6), 333--342 (2017)

\bibitem{chu2019utility}
Chu, L.C., Park, S., Kawamoto, S., Fouladi, D.F., Shayesteh, S., Zinreich, E.S., Graves, J.S., Horton, K.M., Hruban, R.H., Yuille, A.L., et~al.: Utility of ct radiomics features in differentiation of pancreatic ductal adenocarcinoma from normal pancreatic tissue.
\newblock American Journal of Roentgenology \textbf{213}(2), 349--357 (2019)

\bibitem{codella2018skin}
Codella, N.C., Gutman, D., Celebi, M.E., Helba, B., Marchetti, M.A., Dusza, S.W., Kalloo, A., Liopyris, K., Mishra, N., Kittler, H., et~al.: Skin lesion analysis toward melanoma detection: A challenge at the 2017 international symposium on biomedical imaging (isbi), hosted by the international skin imaging collaboration (isic).
\newblock In: 2018 IEEE 15th international symposium on biomedical imaging (ISBI 2018), pp. 168--172. IEEE (2018)

\bibitem{davis2000tumors}
Davis, G.B., Blanchard, D.K., Hatch~III, G.F., Wertheimer-Hatch, L., Hatch, K.F., Foster~Jr, R.S., Skandalakis, J.E.: Tumors of the stomach.
\newblock World journal of surgery \textbf{24}(4), 412--420 (2000)

\bibitem{du2023boosting}
Du, S., Wang, X., Lu, Y., Zhou, Y., Zhang, S., Yuille, A., Li, K., Zhou, Z.: Boosting dermatoscopic lesion segmentation via diffusion models with visual and textual prompts.
\newblock arXiv preprint arXiv:2310.02906  (2023)

\bibitem{dunnick2016renal}
Dunnick, N.R.: Renal cell carcinoma: staging and surveillance.
\newblock Abdominal Radiology \textbf{41}, 1079--1085 (2016)

\bibitem{edge2010american}
Edge, S.B., Compton, C.C.: The american joint committee on cancer: the 7th edition of the ajcc cancer staging manual and the future of tnm.
\newblock Annals of surgical oncology \textbf{17}(6), 1471--1474 (2010)

\bibitem{elbanna2020imaging}
Elbanna, K.Y., Jang, H.J., Kim, T.K.: Imaging diagnosis and staging of pancreatic ductal adenocarcinoma: a comprehensive review.
\newblock Insights into imaging \textbf{11}(1), 1--13 (2020)

\bibitem{engelhardt2019flexible}
Engelhardt, S., Sauerzapf, S., Preim, B., Karck, M., Wolf, I., De~Simone, R.: Flexible and comprehensive patient-specific mitral valve silicone models with chordae tendineae made from 3d-printable molds.
\newblock International journal of computer assisted radiology and surgery \textbf{14}, 1177--1186 (2019)

\bibitem{feng2020parts2whole}
Feng, R., Zhou, Z., Gotway, M.B., Liang, J.: Parts2whole: Self-supervised contrastive learning via reconstruction.
\newblock In: Domain Adaptation and Representation Transfer, and Distributed and Collaborative Learning, pp. 85--95. Springer (2020)

\bibitem{fowler2021pathologic}
Fowler, K.J., Burgoyne, A., Fraum, T.J., Hosseini, M., Ichikawa, S., Kim, S., Kitao, A., Lee, J.M., Paradis, V., Taouli, B., et~al.: Pathologic, molecular, and prognostic radiologic features of hepatocellular carcinoma.
\newblock Radiographics \textbf{41}(6), 1611--1631 (2021)

\bibitem{gao2023synthetic}
Gao, C., Killeen, B.D., Hu, Y., Grupp, R.B., Taylor, R.H., Armand, M., Unberath, M.: Synthetic data accelerates the development of generalizable learning-based algorithms for x-ray image analysis.
\newblock Nature Machine Intelligence \textbf{5}(3), 294--308 (2023)

\bibitem{golias2004cell}
Golias, C., Charalabopoulos, A., Charalabopoulos, K.: Cell proliferation and cell cycle control: a mini review.
\newblock International journal of clinical practice \textbf{58}(12), 1134--1141 (2004)

\bibitem{goodfellow2014generative}
Goodfellow, I., Pouget-Abadie, J., Mirza, M., Xu, B., Warde-Farley, D., Ozair, S., Courville, A., Bengio, Y.: Generative adversarial nets.
\newblock Advances in neural information processing systems \textbf{27} (2014)

\bibitem{goodfellow2020generative}
Goodfellow, I., Pouget-Abadie, J., Mirza, M., Xu, B., Warde-Farley, D., Ozair, S., Courville, A., Bengio, Y.: Generative adversarial networks.
\newblock Communications of the ACM \textbf{63}(11), 139--144 (2020)

\bibitem{griswold1975colon}
Griswold, D., Corbett, T.H.: A colon tumor model for anticancer agent evaluation.
\newblock Cancer \textbf{36}(S6), 2441--2444 (1975)

\bibitem{haghighi2020learning}
Haghighi, F., Hosseinzadeh~Taher, M.R., Zhou, Z., Gotway, M.B., Liang, J.: Learning semantics-enriched representation via self-discovery, self-classification, and self-restoration.
\newblock In: International Conference on Medical Image Computing and Computer-Assisted Intervention, pp. 137--147. Springer (2020).
\newblock \urlprefix\url{https://github.com/fhaghighi/SemanticGenesis}

\bibitem{haghighi2021transferable}
Haghighi, F., Taher, M.R.H., Zhou, Z., Gotway, M.B., Liang, J.: Transferable visual words: Exploiting the semantics of anatomical patterns for self-supervised learning.
\newblock IEEE Transactions on Medical Imaging  (2021).
\newblock \urlprefix\url{https://github.com/fhaghighi/SemanticGenesis}

\bibitem{hamamci2023generatect}
Hamamci, I.E., Er, S., Simsar, E., Tezcan, A., Simsek, A.G., Almas, F., Esirgun, S.N., Reynaud, H., Pati, S., Bluethgen, C., et~al.: Generatect: text-guided 3d chest ct generation.
\newblock In: Proceedings of the European Conference on Computer Vision (ECCV) (2024)

\bibitem{han2019infinite}
Han, C., Rundo, L., Araki, R., Furukawa, Y., Mauri, G., Nakayama, H., Hayashi, H.: Infinite brain mr images: Pggan-based data augmentation for tumor detection.
\newblock In: Neural approaches to dynamics of signal exchanges, pp. 291--303. Springer (2019)

\bibitem{hanahan2000hallmarks}
Hanahan, D., Weinberg, R.A.: The hallmarks of cancer.
\newblock cell \textbf{100}(1), 57--70 (2000)

\bibitem{harpold2007evolution}
Harpold, H.L., Alvord~Jr, E.C., Swanson, K.R.: The evolution of mathematical modeling of glioma proliferation and invasion.
\newblock Journal of Neuropathology \& Experimental Neurology \textbf{66}(1), 1--9 (2007)

\bibitem{heath1998current}
Heath, M., Bowyer, K., Kopans, D., Kegelmeyer~Jr, P., Moore, R., Chang, K., Munishkumaran, S.: Current status of the digital database for screening mammography.
\newblock In: Digital Mammography: Nijmegen, 1998, pp. 457--460. Springer (1998)

\bibitem{heller2021state}
Heller, N., Isensee, F., Maier-Hein, K.H., Hou, X., Xie, C., Li, F., Nan, Y., Mu, G., Lin, Z., Han, M., et~al.: The state of the art in kidney and kidney tumor segmentation in contrast-enhanced ct imaging: Results of the kits19 challenge.
\newblock Medical image analysis \textbf{67}, 101821 (2021)

\bibitem{hiraoka2010tumour}
Hiraoka, N., Ino, Y., Sekine, S., Tsuda, H., Shimada, K., Kosuge, T., Zavada, J., Yoshida, M., Yamada, K., Koyama, T., et~al.: Tumour necrosis is a postoperative prognostic marker for pancreatic cancer patients with a high interobserver reproducibility in histological evaluation.
\newblock British journal of cancer \textbf{103}(7), 1057--1065 (2010)

\bibitem{ho2020denoising}
Ho, J., Jain, A., Abbeel, P.: Denoising diffusion probabilistic models.
\newblock Advances in Neural Information Processing Systems \textbf{33}, 6840--6851 (2020)

\bibitem{hou2019robust}
Hou, L., Agarwal, A., Samaras, D., Kurc, T.M., Gupta, R.R., Saltz, J.H.: Robust histopathology image analysis: To label or to synthesize?
\newblock In: Proceedings of the IEEE/CVF conference on computer vision and pattern recognition, pp. 8533--8542 (2019)

\bibitem{hu2023label}
Hu, Q., Chen, Y., Xiao, J., Sun, S., Chen, J., Yuille, A.L., Zhou, Z.: Label-free liver tumor segmentation.
\newblock In: IEEE/CVF Conference on Computer Vision and Pattern Recognition, pp. 7422--7432 (2023).
\newblock \urlprefix\url{https://github.com/MrGiovanni/SyntheticTumors}

\bibitem{hu2022synthetic}
Hu, Q., Xiao, J., Chen, Y., Sun, S., Chen, J.N., Yuille, A., Zhou, Z.: Synthetic tumors make ai segment tumors better.
\newblock NeurIPS Workshop on Medical Imaging meets NeurIPS  (2022).
\newblock \urlprefix\url{https://github.com/MrGiovanni/SyntheticTumors}

\bibitem{hu2023synthetic}
Hu, Q., Yuille, A., Zhou, Z.: Synthetic data as validation.
\newblock arXiv preprint arXiv:2310.16052  (2023).
\newblock \urlprefix\url{https://github.com/MrGiovanni/SyntheticValidation}

\bibitem{iyer2004imaging}
Iyer, R., Dubrow, R.: Imaging of esophageal cancer.
\newblock Cancer Imaging \textbf{4}(2), 125 (2004)

\bibitem{jaipuria2020deflating}
Jaipuria, N., Zhang, X., Bhasin, R., Arafa, M., Chakravarty, P., Shrivastava, S., Manglani, S., Murali, V.N.: Deflating dataset bias using synthetic data augmentation.
\newblock In: Proceedings of the IEEE/CVF Conference on Computer Vision and Pattern Recognition Workshops, pp. 772--773 (2020)

\bibitem{jiang2020covid}
Jiang, Y., Chen, H., Loew, M., Ko, H.: Covid-19 ct image synthesis with a conditional generative adversarial network.
\newblock IEEE Journal of Biomedical and Health Informatics \textbf{25}(2), 441--452 (2020)

\bibitem{jin2018ct}
Jin, D., Xu, Z., Tang, Y., Harrison, A.P., Mollura, D.J.: Ct-realistic lung nodule simulation from 3d conditional generative adversarial networks for robust lung segmentation.
\newblock In: Medical Image Computing and Computer Assisted Intervention, pp. 732--740. Springer (2018)

\bibitem{jin2021free}
Jin, Q., Cui, H., Sun, C., Meng, Z., Su, R.: Free-form tumor synthesis in computed tomography images via richer generative adversarial network.
\newblock Knowledge-Based Systems \textbf{218}, 106753 (2021)

\bibitem{kang2023label}
Kang, M., Li, B., Zhu, Z., Lu, Y., Fishman, E.K., Yuille, A., Zhou, Z.: Label-assemble: Leveraging multiple datasets with partial labels.
\newblock In: IEEE International Symposium on Biomedical Imaging, pp. 1--5. IEEE (2023).
\newblock \urlprefix\url{https://github.com/MrGiovanni/LabelAssemble}

\bibitem{kang2011cyst}
Kang, M.J., Jang, J.Y., Kim, S.J., Lee, K.B., Ryu, J.K., Kim, Y.T., Yoon, Y.B., Kim, S.W.: Cyst growth rate predicts malignancy in patients with branch duct intraductal papillary mucinous neoplasms.
\newblock Clinical Gastroenterology and Hepatology \textbf{9}(1), 87--93 (2011)

\bibitem{khader2023denoising}
Khader, F., M{\"u}ller-Franzes, G., Tayebi~Arasteh, S., Han, T., Haarburger, C., Schulze-Hagen, M., Schad, P., Engelhardt, S., Bae{\ss}ler, B., Foersch, S., et~al.: Denoising diffusion probabilistic models for 3d medical image generation.
\newblock Scientific Reports \textbf{13}(1), 7303 (2023)

\bibitem{kingma2013auto}
Kingma, D.P., Welling, M.: Auto-encoding variational bayes.
\newblock arXiv preprint arXiv:1312.6114  (2013)

\bibitem{kuijf2019standardized}
Kuijf, H.J., Biesbroek, J.M., De~Bresser, J., Heinen, R., Andermatt, S., Bento, M., Berseth, M., Belyaev, M., Cardoso, M.J., Casamitjana, A., et~al.: Standardized assessment of automatic segmentation of white matter hyperintensities and results of the wmh segmentation challenge.
\newblock IEEE transactions on medical imaging \textbf{38}(11), 2556--2568 (2019)

\bibitem{kumar2017dataset}
Kumar, N., Verma, R., Sharma, S., Bhargava, S., Vahadane, A., Sethi, A.: A dataset and a technique for generalized nuclear segmentation for computational pathology.
\newblock IEEE transactions on medical imaging \textbf{36}(7), 1550--1560 (2017)

\bibitem{kumar2017robbins}
Kumar, V., Abbas, A., Aster, J.C.: Robbins basic pathology.
\newblock Elsevier Health Sciences (2017)

\bibitem{laeseke2015combining}
Laeseke, P.F., Chen, R., Jeffrey, R.B., Brentnall, T.A., Willmann, J.K.: Combining in vitro diagnostics with in vivo imaging for earlier detection of pancreatic ductal adenocarcinoma: challenges and solutions.
\newblock Radiology \textbf{277}(3), 644--661 (2015)

\bibitem{lahouel2020revisiting}
Lahouel, K., Younes, L., Danilova, L., Giardiello, F.M., Hruban, R.H., Groopman, J., Kinzler, K.W., Vogelstein, B., Geman, D., Tomasetti, C.: Revisiting the tumorigenesis timeline with a data-driven generative model.
\newblock Proceedings of the National Academy of Sciences \textbf{117}(2), 857--864 (2020)

\bibitem{lai2024pixel}
Lai, Y., Chen, X., Wang, A., Yuille, A., Zhou, Z.: From pixel to cancer: Cellular automata in computed tomography.
\newblock arXiv preprint arXiv:2403.06459  (2024).
\newblock \urlprefix\url{https://github.com/MrGiovanni/Pixel2Cancer}

\bibitem{lee2004triple}
Lee, K., O'Malley, M., Haider, M., Hanbidge, A.: Triple-phase mdct of hepatocellular carcinoma.
\newblock American Journal of Roentgenology \textbf{182}(3), 643--649 (2004)

\bibitem{leveridge2010imaging}
Leveridge, M.J., Bostrom, P.J., Koulouris, G., Finelli, A., Lawrentschuk, N.: Imaging renal cell carcinoma with ultrasonography, ct and mri.
\newblock Nature Reviews Urology \textbf{7}(6), 311--325 (2010)

\bibitem{lewin1996tumors}
Lewin, K.J., Appelman, H.D., et~al.: Tumors of the esophagus and stomach.
\newblock American Registry of Pathology (1996)

\bibitem{li2023early}
Li, B., Chou, Y.C., Sun, S., Qiao, H., Yuille, A., Zhou, Z.: Early detection and localization of pancreatic cancer by label-free tumor synthesis.
\newblock MICCAI Workshop on Big Task Small Data, 1001-AI  (2023).
\newblock \urlprefix\url{https://github.com/MrGiovanni/SyntheticTumors}

\bibitem{li2024endora}
Li, C., Liu, H., Liu, Y., Feng, B.Y., Li, W., Liu, X., Chen, Z., Shao, J., Yuan, Y.: Endora: Video generation models as endoscopy simulators.
\newblock In: International Conference on Medical Image Computing and Computer-Assisted Intervention (2024)

\bibitem{li2024abdomenatlas}
Li, W., Qu, C., Chen, X., Bassi, P.R., Shi, Y., Lai, Y., Yu, Q., Xue, H., Chen, Y., Lin, X., et~al.: Abdomenatlas: A large-scale, detailed-annotated, \& multi-center dataset for efficient transfer learning and open algorithmic benchmarking.
\newblock Medical Image Analysis p. 103285 (2024).
\newblock \urlprefix\url{https://github.com/MrGiovanni/AbdomenAtlas}

\bibitem{li2024well}
Li, W., Yuille, A., Zhou, Z.: How well do supervised models transfer to 3d image segmentation?
\newblock In: International Conference on Learning Representations (2024).
\newblock \urlprefix\url{https://github.com/MrGiovanni/SuPreM}

\bibitem{liu2023clip}
Liu, J., Zhang, Y., Chen, J.N., Xiao, J., Lu, Y., A~Landman, B., Yuan, Y., Yuille, A., Tang, Y., Zhou, Z.: Clip-driven universal model for organ segmentation and tumor detection.
\newblock In: Proceedings of the IEEE/CVF International Conference on Computer Vision, pp. 21152--21164 (2023).
\newblock \urlprefix\url{https://github.com/ljwztc/CLIP-Driven-Universal-Model}

\bibitem{liu2024universal}
Liu, J., Zhang, Y., Wang, K., Yavuz, M.C., Chen, X., Yuan, Y., Li, H., Yang, Y., Yuille, A., Tang, Y., et~al.: Universal and extensible language-vision models for organ segmentation and tumor detection from abdominal computed tomography.
\newblock Medical Image Analysis p. 103226 (2024).
\newblock \urlprefix\url{https://github.com/ljwztc/CLIP-Driven-Universal-Model}

\bibitem{luo2022adaptive}
Luo, Y., Zhou, L., Zhan, B., Fei, Y., Zhou, J., Wang, Y., Shen, D.: Adaptive rectification based adversarial network with spectrum constraint for high-quality pet image synthesis.
\newblock Medical Image Analysis \textbf{77}, 102335 (2022)

\bibitem{lyu2022learning}
Lyu, F., Ye, M., Ma, A.J., Yip, T.C.F., Wong, G.L.H., Yuen, P.C.: Learning from synthetic ct images via test-time training for liver tumor segmentation.
\newblock IEEE transactions on medical imaging \textbf{41}(9), 2510--2520 (2022)

\bibitem{m2021use}
M.~Cunha, G., Fowler, K.J., Roudenko, A., Taouli, B., Fung, A.W., Elsayes, K.M., Marks, R.M., Cruite, I., Horvat, N., Chernyak, V., et~al.: How to use li-rads to report liver ct and mri observations.
\newblock RadioGraphics \textbf{41}(5), 1352--1367 (2021)

\bibitem{menze2014multimodal}
Menze, B.H., Jakab, A., Bauer, S., Kalpathy-Cramer, J., Farahani, K., Kirby, J., Burren, Y., Porz, N., Slotboom, J., Wiest, R., et~al.: The multimodal brain tumor image segmentation benchmark (brats).
\newblock IEEE transactions on medical imaging \textbf{34}(10), 1993--2024 (2014)

\bibitem{mukherkjee2022brain}
Mukherkjee, D., Saha, P., Kaplun, D., Sinitca, A., Sarkar, R.: Brain tumor image generation using an aggregation of gan models with style transfer.
\newblock Scientific Reports \textbf{12}(1), 9141 (2022)

\bibitem{nerad2016diagnostic}
Nerad, E., Lahaye, M.J., Maas, M., Nelemans, P., Bakers, F.C., Beets, G.L., Beets-Tan, R.G.: Diagnostic accuracy of ct for local staging of colon cancer: a systematic review and meta-analysis.
\newblock American Journal of Roentgenology \textbf{207}(5), 984--995 (2016)

\bibitem{orbes2019multi}
Orbes-Arteaga, M., Varsavsky, T., Sudre, C.H., Eaton-Rosen, Z., Haddow, L.J., S{\o}rensen, L., Nielsen, M., Pai, A., Ourselin, S., Modat, M., et~al.: Multi-domain adaptation in brain mri through paired consistency and adversarial learning.
\newblock In: Domain Adaptation and Representation Transfer and Medical Image Learning with Less Labels and Imperfect Data, pp. 54--62. Springer (2019)

\bibitem{ozbey2023unsupervised}
{\"O}zbey, M., Dalmaz, O., Dar, S.U., Bedel, H.A., {\"O}zturk, {\c{S}}., G{\"u}ng{\"o}r, A., {\c{C}}ukur, T.: Unsupervised medical image translation with adversarial diffusion models.
\newblock IEEE Transactions on Medical Imaging  (2023)

\bibitem{park2021generative}
Park, J.E., Eun, D., Kim, H.S., Lee, D.H., Jang, R.W., Kim, N.: Generative adversarial network for glioblastoma ensures morphologic variations and improves diagnostic model for isocitrate dehydrogenase mutant type.
\newblock Scientific reports \textbf{11}(1), 9912 (2021)

\bibitem{park2020annotated}
Park, S., Chu, L., Fishman, E., Yuille, A., Vogelstein, B., Kinzler, K., Horton, K., Hruban, R., Zinreich, E., Fouladi, D.F., et~al.: Annotated normal ct data of the abdomen for deep learning: Challenges and strategies for implementation.
\newblock Diagnostic and interventional imaging \textbf{101}(1), 35--44 (2020)

\bibitem{qu2023annotating}
Qu, C., Zhang, T., Qiao, H., Liu, J., Tang, Y., Yuille, A., Zhou, Z.: Abdomenatlas-8k: Annotating 8,000 abdominal ct volumes for multi-organ segmentation in three weeks.
\newblock Conference on Neural Information Processing Systems  (2023).
\newblock \urlprefix\url{https://github.com/MrGiovanni/AbdomenAtlas}

\bibitem{qu2024abdomenatlas}
Qu, C., Zhang, T., Qiao, H., Tang, Y., Yuille, A.L., Zhou, Z., et~al.: Abdomenatlas-8k: Annotating 8,000 ct volumes for multi-organ segmentation in three weeks.
\newblock Advances in Neural Information Processing Systems \textbf{36} (2024)

\bibitem{ramesh2022hierarchical}
Ramesh, A., Dhariwal, P., Nichol, A., Chu, C., Chen, M.: Hierarchical text-conditional image generation with clip latents.
\newblock arXiv preprint arXiv:2204.06125 \textbf{1}(2), 3 (2022)

\bibitem{reynolds2015infiltrative}
Reynolds, A.R., Furlan, A., Fetzer, D.T., Sasatomi, E., Borhani, A.A., Heller, M.T., Tublin, M.E.: Infiltrative hepatocellular carcinoma: what radiologists need to know.
\newblock Radiographics \textbf{35}(2), 371--386 (2015)

\bibitem{roth2022towards}
Roth, K., Pemula, L., Zepeda, J., Sch{\"o}lkopf, B., Brox, T., Gehler, P.: Towards total recall in industrial anomaly detection.
\newblock In: Proceedings of the IEEE/CVF Conference on Computer Vision and Pattern Recognition, pp. 14318--14328 (2022)

\bibitem{saar2008radiological}
Saar, B., Kellner-Weldon, F.: Radiological diagnosis of hepatocellular carcinoma.
\newblock Liver International \textbf{28}(2), 189--199 (2008)

\bibitem{sahani2016abdominal}
Sahani, D.V., Samir, A.E.: Abdominal Imaging E-Book: Expert Radiology Series.
\newblock Elsevier Health Sciences (2016)

\bibitem{schlegl2019f}
Schlegl, T., Seeb{\"o}ck, P., Waldstein, S.M., Langs, G., Schmidt-Erfurth, U.: f-anogan: Fast unsupervised anomaly detection with generative adversarial networks.
\newblock Medical Image Analysis  (2019)

\bibitem{setio2017validation}
Setio, A.A.A., Traverso, A., De~Bel, T., Berens, M.S., van~den Bogaard, C., Cerello, P., Chen, H., Dou, Q., Fantacci, M.E., Geurts, B., et~al.: Validation, comparison, and combination of algorithms for automatic detection of pulmonary nodules in computed tomography images: the luna16 challenge.
\newblock Medical image analysis \textbf{42}, 1--13 (2017)

\bibitem{sharan2021mutually}
Sharan, L., Romano, G., Koehler, S., Kelm, H., Karck, M., De~Simone, R., Engelhardt, S.: Mutually improved endoscopic image synthesis and landmark detection in unpaired image-to-image translation.
\newblock IEEE Journal of Biomedical and Health Informatics \textbf{26}(1), 127--138 (2021)

\bibitem{silverman2012oncologic}
Silverman, P.M.: Oncologic imaging: a multidisciplinary approach.
\newblock Elsevier Health Sciences (2012)

\bibitem{skarin2015atlas}
Skarin, A.T.: Atlas of Diagnostic Oncology E-Book.
\newblock Elsevier Health Sciences (2015)

\bibitem{smaldone2012small}
Smaldone, M.C., Kutikov, A., Egleston, B.L., Canter, D.J., Viterbo, R., Chen, D.Y., Jewett, M.A., Greenberg, R.E., Uzzo, R.G.: Small renal masses progressing to metastases under active surveillance: a systematic review and pooled analysis.
\newblock Cancer \textbf{118}(4), 997--1006 (2012)

\bibitem{soga2005early}
Soga, J.: Early-stage carcinoids of the gastrointestinal tract: an analysis of 1914 reported cases.
\newblock Cancer: Interdisciplinary International Journal of the American Cancer Society \textbf{103}(8), 1587--1595 (2005)

\bibitem{stout1953tumors}
Stout, A.P.: Tumors of the stomach.
\newblock Armed Forces Institute of Pathology (1953)

\bibitem{stout1957tumors}
Stout, A.P., Lattes, R.: Tumors of the esophagus.
\newblock Armed Forces Institute of Pathology (1957)

\bibitem{tanase2015complexity}
Tanase, M., Waliszewski, P.: On complexity and homogeneity measures in predicting biological aggressiveness of prostate cancer; implication of the cellular automata model of tumor growth.
\newblock Journal of surgical oncology \textbf{112}(8), 791--801 (2015)

\bibitem{tang2024efficient}
Tang, Y., Liu, J., Zhou, Z., Yu, X., Huo, Y.: Efficient 3d representation learning for medical image analysis.
\newblock World Scientific Annual Review of Artificial Intelligence  (2024)

\bibitem{teixeira2018generating}
Teixeira, B., Singh, V., Chen, T., Ma, K., Tamersoy, B., Wu, Y., Balashova, E., Comaniciu, D.: Generating synthetic x-ray images of a person from the surface geometry.
\newblock In: Proceedings of the IEEE conference on computer vision and pattern recognition, pp. 9059--9067 (2018)

\bibitem{turing2009computing}
Turing, A.M.: Computing machinery and intelligence.
\newblock Springer (2009)

\bibitem{van2017computational}
Van~Griethuysen, J.J., Fedorov, A., Parmar, C., Hosny, A., Aucoin, N., Narayan, V., Beets-Tan, R.G., Fillion-Robin, J.C., Pieper, S., Aerts, H.J.: Computational radiomics system to decode the radiographic phenotype.
\newblock Cancer research \textbf{77}(21), e104--e107 (2017)

\bibitem{veeling2018rotation}
Veeling, B.S., Linmans, J., Winkens, J., Cohen, T., Welling, M.: Rotation equivariant cnns for digital pathology.
\newblock In: Medical Image Computing and Computer Assisted Intervention, pp. 210--218. Springer (2018)

\bibitem{waheed2020covidgan}
Waheed, A., Goyal, M., Gupta, D., Khanna, A., Al-Turjman, F., Pinheiro, P.R.: Covidgan: data augmentation using auxiliary classifier gan for improved covid-19 detection.
\newblock Ieee Access \textbf{8}, 91916--91923 (2020)

\bibitem{wang2017comparison}
Wang, H., Zhou, Z., Li, Y., Chen, Z., Lu, P., Wang, W., Liu, W., Yu, L.: Comparison of machine learning methods for classifying mediastinal lymph node metastasis of non-small cell lung cancer from 18f-fdg pet/ct images.
\newblock EJNMMI research \textbf{7}(1), 1--11 (2017)

\bibitem{wang2018locality}
Wang, Y., Zhou, L., Wang, L., Yu, B., Zu, C., Lalush, D.S., Lin, W., Wu, X., Zhou, J., Shen, D.: Locality adaptive multi-modality gans for high-quality pet image synthesis.
\newblock In: Medical Image Computing and Computer Assisted Intervention, pp. 329--337. Springer (2018)

\bibitem{pmid29668296}
Wang, Z.J., Westphalen, A.C., Zagoria, R.J.: {{C}{T} and {M}{R}{I} of small renal masses}.
\newblock Br J Radiol \textbf{91}(1087), 20180131 (2018)

\bibitem{wei2022pancreatic}
Wei, Z., Chen, Y., Guan, Q., Hu, H., Zhou, Q., Li, Z., Xu, X., Frangi, A., Chen, F.: Pancreatic image augmentation based on local region texture synthesis for tumor segmentation.
\newblock In: International Conference on Artificial Neural Networks, pp. 419--431. Springer (2022)

\bibitem{wong2015pancreatic}
Wong, K.C., Summers, R.M., Kebebew, E., Yao, J.: Pancreatic tumor growth prediction with multiplicative growth and image-derived motion.
\newblock In: International Conference on Information Processing in Medical Imaging, pp. 501--513. Springer (2015)

\bibitem{wu2018conditional}
Wu, E., Wu, K., Cox, D., Lotter, W.: Conditional infilling gans for data augmentation in mammogram classification.
\newblock In: Image Analysis for Moving Organ, Breast, and Thoracic Images: Third International Workshop, pp. 98--106. Springer (2018)

\bibitem{xia2022felix}
Xia, Y., Yu, Q., Chu, L., Kawamoto, S., Park, S., Liu, F., Chen, J., Zhu, Z., Li, B., Zhou, Z., et~al.: The felix project: Deep networks to detect pancreatic neoplasms.
\newblock medRxiv  (2022)

\bibitem{xiao2022catenorm}
Xiao, J., Yu, L., Zhou, Z., Bai, Y., Xing, L., Yuille, A., Zhou, Y.: Catenorm: Categorical normalization for robust medical image segmentation.
\newblock In: MICCAI Workshop on Domain Adaptation and Representation Transfer, pp. 129--146. Springer (2022).
\newblock \urlprefix\url{https://github.com/lambert-x/CateNorm}

\bibitem{xue2021selective}
Xue, Y., Ye, J., Zhou, Q., Long, L.R., Antani, S., Xue, Z., Cornwell, C., Zaino, R., Cheng, K.C., Huang, X.: Selective synthetic augmentation with histogan for improved histopathology image classification.
\newblock Medical image analysis \textbf{67}, 101816 (2021)

\bibitem{yan2020mri}
Yan, W., Huang, L., Xia, L., Gu, S., Yan, F., Wang, Y., Tao, Q.: Mri manufacturer shift and adaptation: increasing the generalizability of deep learning segmentation for mr images acquired with different scanners.
\newblock Radiology: Artificial Intelligence \textbf{2}(4), e190195 (2020)

\bibitem{yang2020covid}
Yang, X., He, X., Zhao, J., Zhang, Y., Zhang, S., Xie, P.: Covid-ct-dataset: a ct scan dataset about covid-19.
\newblock arXiv preprint arXiv:2003.13865  (2020)

\bibitem{yao2021label}
Yao, Q., Xiao, L., Liu, P., Zhou, S.K.: Label-free segmentation of covid-19 lesions in lung ct.
\newblock IEEE Transactions on Medical Imaging  (2021)

\bibitem{yao2022unsupervised}
Yao, Y., Liu, F., Zhou, Z., Wang, Y., Shen, W., Yuille, A., Lu, Y.: Unsupervised domain adaptation through shape modeling for medical image segmentation.
\newblock arXiv preprint arXiv:2207.02529  (2022)

\bibitem{yee2021tumor}
Yee, P.P., Li, W.: Tumor necrosis: A synergistic consequence of metabolic stress and inflammation.
\newblock Bioessays \textbf{43}(7), 2100029 (2021)

\bibitem{yoon2022colonoscopic}
Yoon, D., Kong, H.J., Kim, B.S., Cho, W.S., Lee, J.C., Cho, M., Lim, M.H., Yang, S.Y., Lim, S.H., Lee, J., et~al.: Colonoscopic image synthesis with generative adversarial network for enhanced detection of sessile serrated lesions using convolutional neural network.
\newblock Scientific reports \textbf{12}(1), 261 (2022)

\bibitem{yu20183d}
Yu, B., Zhou, L., Wang, L., Fripp, J., Bourgeat, P.: 3d cgan based cross-modality mr image synthesis for brain tumor segmentation.
\newblock In: 2018 IEEE 15th international symposium on biomedical imaging (ISBI 2018), pp. 626--630. IEEE (2018)

\bibitem{zhang2023continual}
Zhang, Y., Li, X., Chen, H., Yuille, A.L., Liu, Y., Zhou, Z.: Continual learning for abdominal multi-organ and tumor segmentation.
\newblock In: International Conference on Medical Image Computing and Computer-Assisted Intervention, pp. 35--45. Springer (2023).
\newblock \urlprefix\url{https://github.com/MrGiovanni/ContinualLearning}

\bibitem{zhao2020tripartite}
Zhao, J., Li, D., Kassam, Z., Howey, J., Chong, J., Chen, B., Li, S.: Tripartite-gan: Synthesizing liver contrast-enhanced mri to improve tumor detection.
\newblock Medical image analysis \textbf{63}, 101667 (2020)

\bibitem{zhou2021towards}
Zhou, Z.: Towards annotation-efficient deep learning for computer-aided diagnosis.
\newblock Ph.D. thesis, Arizona State University (2021).
\newblock \urlprefix\url{https://github.com/MrGiovanni/Dissertation}

\bibitem{zhou2022interpreting}
Zhou, Z., Gotway, M.B., Liang, J.: Interpreting medical images.
\newblock In: Intelligent Systems in Medicine and Health, pp. 343--371. Springer (2022)

\bibitem{zhou2019models}
Zhou, Z., Sodha, V., Siddiquee, M.M.R., Feng, R., Tajbakhsh, N., Gotway, M.B., Liang, J.: Models genesis: Generic autodidactic models for 3d medical image analysis.
\newblock In: International Conference on Medical Image Computing and Computer-Assisted Intervention, pp. 384--393. Springer (2019).
\newblock \urlprefix\url{https://github.com/MrGiovanni/ModelsGenesis}

\bibitem{zhu2022assembling}
Zhu, Z., Kang, M., Yuille, A., Zhou, Z.: Assembling and exploiting large-scale existing labels of common thorax diseases for improved covid-19 classification using chest radiographs.
\newblock In: Radiological Society of North America (RSNA) (2022).
\newblock \urlprefix\url{https://github.com/MrGiovanni/LabelAssemble}

\end{thebibliography}

\end{document}